\documentclass[11pt]{article}
\usepackage[a4paper,hmarginratio=1:1,
totalwidth=15.5cm,totalheight=22.7cm]{geometry}
\usepackage{bm,epstopdf,epsfig,amsmath,amssymb,amsfonts,colordvi,latexsym,comment,cancel,
verbatim}
\usepackage{graphicx,graphics}
\usepackage[font=md,captionskip=8pt]{subfig}

\usepackage[usenames,dvipsnames]{color}

\renewcommand{\baselinestretch}{1.14}

\newcommand{\cL}{{\cal L}}
\newcommand{\cM}{{\cal M}}

\newcommand{\cO}{{\cal O}}
\newcommand{\cP}{{\cal P}}

\newcommand{\ra}{\rightarrow}
\newcommand{\be}{\begin{equation}}
\newcommand{\ee}{\end{equation}}
\newcommand{\bea}{\begin{eqnarray}}
\newcommand{\eea}{\end{eqnarray}}

\newcommand{\baa}{\begin{array}}
\newcommand{\eaa}{\end{array}}
\long\def\symbolfootnote[#1]#2{\begingroup
\def\thefootnote{\fnsymbol{footnote}}\footnote[#1]{#2}\endgroup}
\begin{document}

\thispagestyle{empty}
\begin{flushright}
CERN-PH-TH/2012-054\\
\today
\end{flushright}

\vspace{2cm}

\begin{center}
{\Large\bf Tuning supersymmetric models at the LHC:

\medskip
A comparative analysis at two-loop level.}
\vspace{1cm}

{\bf D. M. Ghilencea$^{\,a,b}$,  H. M. Lee$^{\,a}$, M. Park$^{\,a,}$\footnote{E-mail: 
 dumitru.ghilencea@cern.ch, hyun.min.lee@cern.ch, myeonghun.park@cern.ch}}

\bigskip
{\small  $^a$ Theory Division, CERN, 1211 Geneva 23, Switzerland.}

{\small $^b$ Theoretical Physics Department, National Institute of Physics}

{\small and Nuclear Engineering Bucharest (IFIN-HH), MG-6 077125 Romania.}
\end{center}

\bigskip
\begin{abstract}\noindent
We provide a comparative study of the fine tuning amount ($\Delta$) at the two-loop 
leading log level in supersymmetric models  commonly used in SUSY searches at the LHC.
These are  the constrained MSSM (CMSSM), non-universal Higgs masses models (NUHM1, NUHM2), 
non-universal gaugino masses model (NUGM) and GUT related gaugino masses models (NUGMd).
Two definitions of the fine tuning are used, the first ($\Delta_{max}$)
measures  maximal fine-tuning wrt individual parameters while the second 
($\Delta_q$) adds their contribution  in ``quadrature''. 
As a direct  {consequence} of
 two theoretical constraints (the EW minimum conditions),
fine tuning ($\Delta_q$) emerges  {at the mathematical level} as a
suppressing  factor  (effective prior) of the averaged likelihood ($\cL$)
under the priors, under the integral of the global  probability of measuring 
the data  (Bayesian evidence $p(D)$).   For each model, there is little 
 difference between  $\Delta_q$, $\Delta_{max}$ in the region 
allowed by the data,  with similar behaviour as functions of the Higgs, gluino, stop
mass or SUSY scale ($m_{susy}=(m_{\tilde t_1} m_{\tilde t_2})^{1/2}$) or 
 dark matter and $g\!-\!2$ constraints. 
The analysis has the advantage that by replacing any of these mass scales 
or constraints by their latest bounds
one easily infers for each model the  value of $\Delta_q$, $\Delta_{max}$ or vice versa.
For all models, minimal fine tuning is achieved for $M_{higgs}$ near $115$  GeV
with a $\Delta_q\approx\Delta_{max}\approx 10$ to $100$ depending on the model,
 and  in the CMSSM  this is actually a global minimum. 
Due to  a strong  ($\approx$ exponential) dependence  of $\Delta$ on 
$M_{higgs}$, for a Higgs mass near  $125$ GeV, the above values of  
$\Delta_q\approx \Delta_{max}$  increase to between $500$ and  $1000$.  
Possible corrections to these values are briefly  discussed.
\end{abstract}

\newpage

\section{Introduction}\label{introduction}

Low energy (TeV-scale) supersymmetry (SUSY) can provide a solution to the hierarchy problem. 
This is done without undue amount of electroweak scale fine tuning ($\Delta$) that is
present in the non-supersymmetric theories like the Standard Model (SM).
A large value of this  $\Delta$ is just another face of 
the hierarchy problem (for a review see \cite{Cassel:2011zd} and 
references therein). However, negative searches for superpartners 
increase the SUSY scale ($m_{susy}$)  which in turn can increase  
$\Delta$. In the extreme case when $m_{susy}$  is very high ($\gg$ TeV) 
one recovers the scenario of non-supersymmetric theories (SM, etc)
 with a  large $\Delta$. In the light of current  {negative}
SUSY searches at the LHC 
it is useful to examine in detail the amount of fine tuning 
that supersymmetric models need, as a test of SUSY as a 
solution to the hierarchy problem. The alternative is to ignore this problem and 
adopt an effective theory  approach,  with a low effective cutoff (few TeV) that, 
unlike SUSY, does not detail the ``new physics'' at/beyond this scale.
Such models usually have a $\Delta$  relative to the TeV scale 
comparable to that of SUSY models  relative to the Planck scale.

While  a small value of $\Delta$ (say less than $100$) is desirable, 
the exact value still accepted for a solution 
to the hierarchy problem  is rather subjective. Even worse, there are 
also different definitions  of $\Delta$ in the literature. Two common definitions are
\smallskip
\bea\label{tun}
\Delta_{max}=\max\big\vert \Delta_{\gamma_i} \big\vert,
\quad
\Delta_q=\Big(\sum \Delta_{\gamma_i}^2\Big)^{1/2},
\quad 
\Delta_{\gamma_i}=\frac{\partial \ln v}{\partial \ln \gamma_i},\quad
\gamma_i=m_0, m_{1/2}, \mu_0, A_0, B_0.
\eea

\smallskip\noindent
$\gamma_i$ are new parameters (of mass dimension 1), 
that SUSY introduces in the model (shown above for the CMSSM).
$\Delta_{max}$  was the first measure used  \cite{Ellis:1986yg}, 
but  $\Delta_q$ is also common. Two definitions for $\Delta$ can 
lead to different  predictions and the absence of a widely accepted 
upper value for it is another problem.
To avoid these issues, we compute both $\Delta_q$ and $\Delta_{max}$
 and compare their implications in generic SUSY models, 
without an upper bound on them (to be fixed by the reader).
This  is one of the main  purposes  of this work.

These measures of fine tuning were  introduced more on physical 
intuition than rigorous mathematical grounds so another  {important}
purpose is to clarify their link with other approaches and find 
technical support.  Both $\Delta$ provide a  local
measure (in the space $\gamma_i$) of the quantum cancellations
and help to decide which phase space points
of a  model are less fine tuned (more probable). When actually 
comparing  models, a more  global measure would be desirable.
Our scan over the whole parameter  space when evaluating $\Delta$'s 
will alleviate this issue. But one question remains: what is the 
relation of $\Delta$  to other (global) measures of the success of SUSY 
in solving the hierarchy problem? 
 {To answer this, consider the Bayesian probability density  $\cP(\gamma_i\vert D)$  
of a point in parameter space $\{\gamma_i\}$ given the data $D$:}
\smallskip
\bea\label{dd}
\cP(\gamma_i\vert D)=\frac{1}{p(D)}\,\cL(D\vert \gamma_i)\,\,p(\gamma_i),\qquad 
p(D)=\int \cL(D\vert \gamma_i)\,p(\gamma_i)d\gamma_i.
\eea

\smallskip\noindent
Here $\cL$ is the likelihood  the parameters $\{\gamma_i\}$ fit the data $D$ and
 $p(D)$ is a  global  normalization factor called  {Bayesian} ``evidence''.
For two models with the same data and  priors $p(\gamma_i)$, 
the ratio of  their $p(D)$ gives their relative  overall probability. 
So a large $p(D)$  is needed to decide that a model is more probable  {than another}.
Then what is the relation between $p(D)$ 
 {and fine-tuning?
As it was observed in  \cite{Cabrera:2008tj}  (see also \cite{AbdusSalam:2009qd}),  when}
 integrating $\cP(\gamma_i\vert D)$ over one parameter of the theory (in this case $\mu_0$), 
following an {\it experimental} constraint (on $m_Z$), 
there is an emergent {\it effective} prior  
$p_{eff}\!\approx\! 1/\Delta_{\mu_0}$   
 {which brings in} a fine tuning penalty for points with large\footnote{Note
however that fine tuning wrt $\mu_0$ is not dominant  in CMSSM for higgs mass above 
$\sim 115$ GeV \cite{Cassel:2010px}.}
$\Delta_{\mu_0}\sim \Delta_{max}$ \cite{Cabrera:2008tj}.
These points then  have little contribution
to   $p(D)$ because  $\int \cP \sim \cL\times p_{eff}\sim 
(1/\Delta_{max})\,\,\cL$. A larger $p(D)$ can
then indicate a  preference for points of  lower $\Delta_{max}$, 
and the link of $p(D)$  with fine tuning  {wrt individual parameters}
is apparent. 

We explore this idea further and evaluate  $p(D)$  by  {investigating
the effect} of the {\it theoretical } constraints that received less attention:
we refer to the two minimum conditions of the potential.
 {Further, the above observation  and the need to evaluate $p(D)$
clearly suggests to integrate over {\it all} $\{\gamma_i\}$ parameters (and we shall do so),
and also over nuisance variables,  which  are parameters already present in the SM  
(like Yukawa couplings   \cite{Cabrera:2008tj,nui}).} 
The result is that $p(D)\!\sim \! 1/\Delta_q$ so $\Delta_q$ is actually
preferred by the Bayesian evidence calculation.
Then $p(D)$  receives contributions mostly from points of
small  $\Delta_q$, but this also depends  on the 
priors and $\cL$. To conclude, the inverse of 
$\Delta_q$  acts as an extra, effective prior in  
(\ref{dd}) and is indeed  a  physical quantity with  impact on global $p(D)$. 
 {This clarifies the exact, mathematical link of overall
fine tuning wrt all parameters $\{\gamma_i\}$, to $\cL(D\vert\gamma_i)$
and the Bayesian evidence $p(D)$.}

With  {this technical} motivation,  we then evaluate  the 
fine tuning for generic models,  using both definitions $\Delta_q$ and $\Delta_{max}$; 
this is done in a two-loop leading log numerical analysis that provides the 
state-of-the-art  analysis of the fine tuning in the models considered, 
consistent with current data.  Interestingly, the results we find 
 are little dependent on the definition used for $\Delta$, with $\Delta_q$ and
$\Delta_{max}$ showing  similar behaviour and values.
This is important since it is usually thought that  
 {different} fine  tuning measures 
 {should} give different results.
Our results correspond to a scan over the entire   parameter 
space of the  models (including $\tan\beta$).  
This is an extremely CPU-intensive task, made 
possible by the CERN batch computing service.  The analysis of $\Delta_{max}$, 
$\Delta_q$  is done for the  following models:

a) - CMSSM: the constrained minimal supersymmetric standard model. For a recent 

two-loop  leading log  analysis of this topic see \cite{Cassel:2010px}, and 
for earlier  investigations see \cite{Barbieri:1998uv}.

b) - NUHM1: a CMSSM-like model but with equal Higgs soft masses, different from $m_0$.

c) - NUHM2: as for CMSSM but  with different Higgs soft masses and  different 
from $m_0$.

d) - NUGM: a CMSSM-like model but with non-universal gaugino masses.

e) - NUGMd: a benchmark NUGM model \cite{Horton:2009ed} 
with a GUT relation among gaugino masses. 

For these models our results are presented in a comparative way
with $\Delta_q$, $\Delta_{max}$ as functions of the lightest higgs, gluino, stop mass
or the SUSY scale. Any experimental constraints on 
these can easily be used to identify $\Delta_q$, $\Delta_{max}$ 
for that model.  
On top of these plots various contour lines corresponding to the 
remaining masses, dark matter or  the $g-2$ constraints are shown.  
Such  comparative analysis for different models 
and definitions of $\Delta$ was not done in the past and has the advantage that
it can be  updated by the latest data, without re-doing the 
whole analysis. In particular, for each model we identify 
the corresponding $\Delta$'s  for a Higgs mass of $M_{higgs}\!=\!125\pm 2$  
GeV that seems favoured by Atlas and CMS \cite{atlas2011,CMS2011}. 
We shall see such value requires 
$\Delta_{q}\!\approx\! \Delta_{max}\!\sim\! 500$ to $1000$ depending on the model,
and uncertainties in $\Delta$ are also discussed.
In all cases $\Delta$ is minimal near $M_{higgs}\!\approx\! 115$ GeV.
For ways to have  $M_{higgs}\!\approx\! 125-130$ GeV 
with smaller fine-tuning $\Delta\!\approx \!\cO(10)$ 
in SUSY models see \cite{Cassel:2011zd,Cassel:2009ps}.

In the following Section~\ref{th} shows the link of $\Delta$ to the 
evidence $p(D)$ in models with theoretical constraints.
Numerical results and corresponding plots of $\Delta$ are 
shown in Section~\ref{num}.

\section{Fine tuning, $p(D)$  and the role of theoretical constraints.}
\label{th}

Before our numerical analysis, we re-examine 
the relation between  the Bayesian probability of a point in 
parameter space or the evidence $p(D)$  and the  EW scale  fine tuning, 
in models with  theoretical constraints. Without loss of generality,
we do this for the constrained MSSM (CMSSM).
This analysis extends a previous similar study of this problem of
\cite{Cabrera:2008tj}  (Section 2 in both papers), 
see also \cite{AbdusSalam:2009qd}.  To place this discussion on 
quantitative grounds consider the CMSSM scalar  potential
\medskip
\begin{eqnarray} 
V&=&  m_1^2\,\,\vert H_1\vert^2
+  m_2^2\,\,\vert H_2\vert^2
- (m_3^2\,\,H_1 \cdot H_2+h.c.)\nonumber\\[3pt]
 &&
 ~+~
(\lambda_1/2) \,\vert H_1\vert^4
+(\lambda_2/2) \,\vert H_2\vert^4
+\lambda_3 \,\vert H_1\vert^2\,\vert H_2\vert^2\,
+\lambda_4\,\vert H_1\cdot H_2 \vert^2\nonumber\\[3pt]
 &&
 ~+~
\Big[\,(\lambda_5/2)\,\,(H_1\cdot  H_2)^2
+\lambda_6\,\,\vert H_1\vert^2\, (H_1 \cdot H_2)+
\lambda_7\,\,\vert H_2 \vert^2\,(H_1 \cdot H_2)+h.c.\Big].
\label{2hdm}
\end{eqnarray}

\medskip\noindent
The couplings $\lambda_j$ and the soft masses receive
one- and two-loop corrections that for the MSSM can be  found
in \cite{Martin:1993zk,Carena:1995bx}. 
Let us introduce the notation
\medskip
\begin{eqnarray}
m^2 &\equiv &
 m_1^2 \, \cos^2 \beta +  m_2^2
 \, \sin^2 \beta - m_3^2 \, \sin 2\beta\nonumber\\[3pt]
\lambda &\equiv &\frac{ \lambda_1^{} }{2} \, \cos^4 \beta 
+ \frac{ \lambda_2^{} }{2} \,  \sin^4 \beta 
+ \frac{ \lambda_{345}^{} }{4} \, \sin^2 2\beta 
+ \sin 2\beta \left( \lambda_6^{} \cos^2 \beta 
+ \lambda_7^{} \sin^2 \beta \right)
\label{ml}
\end{eqnarray} 
where  $\lambda_{345}=\lambda_3+\lambda_4+\lambda_5$. 

When testing a model such as the CMSSM, one  imposes two classes of 
constraints: theoretical and experimental. Let us discuss them.
Minimizing this scalar potential leads to two theoretical constraints
given below and their solutions for $v, \tan\beta$
 are the same as those of the eqs $f_1\!=\!f_2\!=\!0$ where $f_1$ and $f_2$ 
are introduced for later convenience:
\bea
v^2+\frac{m^2}{\lambda}=0,
&&\!\!\!
f_1(\gamma_i; v, \beta, y_t, y_b, \cdots)\equiv 
v-\!\Big(\!-\frac{m^2}{\lambda}\Big)^{1/2}\!\!\!\!,
\,\,\,\,
\gamma_i\!=\!\{ m_0, m_{1/2}, \mu_0, A_0, B_0\}
\nonumber\\
2 \lambda \frac{\partial m^2}{\partial \beta}-m^2
\frac{\partial\lambda}{\partial\beta}=0,
&&\!\!\!
f_2(\gamma_i; v, \beta, y_t, y_b, \cdots)\equiv
\tan\beta-\tan\beta_0(\gamma_i,v,y_t,y_b...),
\label{min}
\eea

\medskip\noindent
Here $v=\sqrt{v_1^2+v_2^2}$ is a combination of vev's of $h_1^{0}$, $h_2^0$.
The order of the arguments of $f_{1,2}$ is relevant later, while the dots 
denote other parameters (other Yukawa or gauge couplings,...) present at 
one-loop and beyond, that we ignore in this section only, 
without loss of generality.

As a result of these two constraints, the EW minimum solutions
 for  $v$ and $\tan\beta$   become functions 
of the (mass dimension 1) parameters $\gamma_i$ of the model 
 which for CMSSM are shown above, in a standard notation.
When discussing fine tuning, usually
only the first  constraint in (\ref{min}) is  considered, 
although the second is equally important, as our result for $p(D)$
will show. These constraints fixing $v$, $\tan\beta$
are assumed to be factorized out from the 
likelihood  function $\cL(D\vert \gamma_i)$ and can be imposed by Dirac 
delta functions of arguments:
\medskip
\bea
\label{er}
\delta\,\big( f_1(\gamma_i; v, \beta, y_t, y_b)\big),\,\,\,\,\,
\delta\,\big(f_2(\gamma_i; v, \beta, y_t, y_b)\big), \,\,\,
i=\overline{1,5}.
\eea

There is also a second class of constraints, that comes from the experiment, such 
as  the accurate measurement of the masses of the
top ($m_t$), bottom ($m_b$)  and  $Z$  boson ($m_Z$). 
Given the high accuracy of these measurements, 
one can assume  some Gaussian distributions for the associated priors 
when evaluating the probability density $\cP(\gamma_i\vert D)$ or the evidence $p(D)$.
However,  for a more qualitative analysis and  to  good approximation one can again
implement  these constraints (likelihood) by Dirac delta functions of  suitable arguments
\medskip
\bea\label{exp}
\delta(m_t-m_t^0); \,\,\,\,\delta(m_b-m_b^0); 
\,\,\,\,\delta(m_Z-m_Z^0),
\eea
where $m_t^0, m_b^0, m_Z^0$ are experimental values. 
One can consider similar constraints for $\alpha_{em}$ and $\alpha_3$ 
gauge couplings but for simplicity we do not do that (their implementation is similar).

When testing the SUSY models with a given set of parameters
(such as $\gamma_{1,...5}$  for CMSSM),  one should in principle  marginalize 
(i.e. integrate) the density $\cP(\gamma_i\vert D)$ over the  ``nuisance''  
parameters. Examples of these nuisance parameters are 
those {\it already present} in the Standard Model. 
These are the Yukawa couplings $y_t, y_b,...$ \cite{Cabrera:2008tj,nui} 
which were restricted (in the analysis of this section only)
 to the simpler case of top and bottom 
Yukawa couplings.  Other  parameters to integrate over are the dependent 
parameters: the  vev $v$ and $\tan\beta$  which can 
(in principle) take any value, until fixed by minimization constraints 
 (\ref{min}), (\ref{er}), also (\ref{exp}) for $v$.

To compare two SUSY models, one should compare their  evidence $p(D)$
for similar priors and data $D$.
To compute $p(D)$, one  integrates over all parameters (of the SM and 
those  mentioned above) and over  $\gamma_i$ as well, with 
chosen priors $p(\gamma_i)$. For the CMSSM case, after imposing 
the above constraints  with the corresponding priors, 
one finds
\medskip
\bea
p(D) &=& N
\int  d\gamma_1... d \gamma_5\,\, p(\gamma_1,...\gamma_5)
\,d y_t \, \,d y_b \,\, d v\,\,  d(\tan\beta)\,\,
     p(y_t)\,\, p(y_b)\,\,\, 
\nonumber\\[2pt]
&\times &
\delta(m_Z-m_Z^0)\,\,\delta(m_t -m_t^0)\,\,\delta(m_b-m_b^0)\nonumber\\[4pt]
&\times &
\delta\,\big(f_1(\gamma_i; v, \beta, y_t, y_b)\big)\,\,
\delta\,\big(f_2(\gamma_i; v, \beta, y_t, y_b)\big)\,
\,\,\cL(D\vert \gamma_{1,2,...5}; \beta, v,y_t,y_b).
\label{opop}
\eea

\medskip\noindent
where $\cL(D\vert \gamma_i; \beta,v,y_t,y_b..)$ 
is the likelihood of fitting the given data (D)
with a particular  set of values  $\gamma_i$; i=1,...5, etc; the
priors $p(\gamma_1,...\gamma_5)$ and $p(y_{t,b})$ are not known, 
but logarithmic or flat  priors are common choices for individual parameters.
Regarding the priors  $p(v)$ and $p(\tan\beta)$, 
these are already included and given by the corresponding Dirac delta's shown in 
(\ref{er}), (\ref{exp}). We integrated over $y_t, y_b$  rather than over the corresponding
masses $m_t, m_b$. This is a possible choice, preferable 
because the masses are {\it derived} quantities, 
see discussion in  \cite{Cabrera:2008tj}.
Finally, leaving aside the integral over $\gamma_j$ and $p(\gamma_1,...\gamma_5)$, the
above equation  simply gives the probability density $\cP(\gamma_i\vert D)$.

The important point about eq.(\ref{opop}) is that now all
parameters $\gamma_i, v, \tan\beta, y_t, y_b,\cdots$ that we integrated over
can be regarded as arbitrary, since the constraints that render them 
dependent variables are implemented by the Dirac delta functions associated to the
theoretical and experimental constraints.
$\cL$ is a function of the CMSSM parameters, but also of 
the nuisance parameters ($y_{t,b}$) and $v, \tan\beta$.
Finally $N$ is a normalization constant not  important below.

To evaluate $p(D)$, one uses $m_Z={g \, v}/{2}$, $m_t={y_t\, v\,\sin\beta}/{\sqrt 2}$,
$m_b= {y_t\, v\,\cos\beta}/{\sqrt 2}$
and after performing the  integrals over $y_t, y_b$ and $v$ one finds
\medskip
\bea\label{twodeltas}
p(D)&=&
\frac{8 N}{g\, v_0^2}
\int  
d\gamma_1.... d\gamma_5\,\, p(\gamma_1,...\gamma_5) \,\,
d(\tan\beta) \, \,p\big(\tilde y_t(\beta)\big)\, 
p\big(\tilde y_b(\beta)\big)\csc(2\beta)\,\qquad\,\,
\nonumber\\[3pt]
&\times &
\delta\big[ f_1\big(\gamma_i;\, \beta,\, v_0,
\,\tilde y_t(\beta), \,\tilde y_b(\beta)\big)\big]\,\,
\delta\big[ f_2\big(\gamma_i;\, \beta,\, v_0,
\,\tilde y_t(\beta), \,\tilde y_b(\beta)\big)\big]\,\,
\nonumber\\[6pt]
&\times &
\cL\big( D\vert \gamma_i;\beta,\, v_0,\, \tilde y_t(\beta),\, \tilde y_b(\beta)\big),
\eea

\medskip\noindent
with  $g^2=g_1^2+g_2^2$ where $g_1$ ($g_2$)
is the gauge couplings of U(1)  (SU(2)) and
\bea\label{stt}
v_0 \equiv  {2 m_Z^0}/{g}=246 \textrm{GeV},\qquad 
\tilde y_t(\beta)\equiv \sqrt 2\,m_t^0/(v_0 \sin\beta),\qquad
\tilde y_b(\beta)\equiv \sqrt 2\,m_b^0/(v_0 \cos\beta).
\eea
%
Integrating over\footnote{We use
 $\delta(g(x))=\delta(x-x_0 ) /\vert g^\prime \big\vert_{x=x_0}$ with 
 $g^\prime$ the derivative wrt $x$ evaluated in $x_0$; $x_0$ is the unique root of 
$g(x_0)=0$;  we apply this to a function
 $g(\beta)=f_2(\gamma_i; \beta,v_0, \tilde y_t(\beta), \tilde y_b(\beta)))$ 
for $x\equiv \tan\beta$ with the root $\beta_0=\beta_0(\gamma_i)$.}
$\beta$:
\medskip
\bea
p(D) &=& 
\frac{8 N}{g\, v_0^2}
\int \,\, d\gamma_1.... d\gamma_5\,\,p(\gamma_1,...\gamma_5) \,\,
\Big\{ p\big(\tilde y_t(\beta)\big)\,  p\big(\tilde y_b(\beta))\big) 
\csc (2\beta)\,\,\big[(f_2)_{\beta}^\prime\big]^{-1}\,\,
\nonumber\\
&\times &  
\delta\big[ f_1\big(\gamma_i;\, 
\beta,\, v_0,\,\tilde y_t(\beta), \,\tilde y_b(\beta)\big)\big]\
\cL\big( D\vert \gamma_i;\beta,\, v_0,\, \tilde y_t(\beta),\, \tilde y_b(\beta)\big)
\Big\}_{\beta=\beta_0(\gamma_i)}
\nonumber\\
&=&
\frac{4 N}{g\, v_0^4}
\int_{\cM} \,\, d S_{\gamma}\,\,\gamma_1....\gamma_5\,\,p(\gamma_1,...\gamma_5) 
\Big\{p\big(\tilde y_t(\beta)\big)\, p\big(\tilde y_b(\beta))\big)
\csc(2\beta)\,\, 
\nonumber\\[2pt]
&\times &
 \big[(f_2)_{\beta}^\prime\, 
\vert \nabla_{\gamma_i}\ln \tilde v (\gamma_i; \beta_0(\gamma_i))
 \vert\big]^{-1}\,\,
\cL\big( D\vert \gamma_i;\beta,\, v_0,\, \tilde y_t(\beta),\, \tilde y_b(\beta)\big)
\Big\}_{\beta=\beta_0(\gamma_i)}.
\label{pofD}
\eea

\medskip\noindent
Above ${(f_2)}_{\beta}^\prime$ denotes the partial 
derivative wrt the variable $\tan\beta$ of the function $f_2$ of arguments:
$f_2(\gamma_i; \beta, v_0, \tilde y_t(\beta), \tilde y_b(\beta))$,
where (\ref{stt}) is used. The curly bracket is evaluated at the 
unique root $\beta=\beta_0(\gamma_i)$
 of the second minimum condition in (\ref{min}) of the scalar 
potential: $f_2=0$. Through this condition,  $\beta$ becomes 
 a function of the independent parameters $\gamma_i$, as usual (one can 
eventually trade $\beta_0$ for $B_0$,  as often done, 
but we do not do this here). In the last step we converted the 
integral into a surface integral\footnote{One  uses 
$\int_{R^n} f(z_1,...,z_n)\, 
\delta(g(z_1,...,z_n))\,d z_1.... d z_n\!=\!\int_{S_{n-1}} f(z_1,...z_n)\,d S_{n-1} 
1/\vert\nabla_{z_i} g\vert$ with $z_i\rightarrow\ln \gamma_i$ where $S_{n-1}$ is
defined by $g=0$ and $\nabla$ is in basis $z_i=\ln\gamma_i$.
Another form of (\ref{pofD}) is found by replacing 
$dS$, $\nabla$ by their values in $\{\gamma_i\}$  space (instead of $\{\ln\gamma_i\}$)
and removing the  product $\gamma_1....\gamma_5$ in integral (\ref{pofD}).}
where $\cM$ is the surface defined by the equation 
$f_1=0$  while   $d S_{\gamma}$ is the surface element in 
the parameter space $\{\ln\gamma_i\}$.
Recall that $f_1=0$ is one minimum condition 
which together  with the second one $\beta=\beta_0(\gamma_i)$ (or $f_2=0$)
 control the value of $p(D)$.
A notation was used $\nabla_{\gamma_i}
 f_1(\gamma_i; \beta, v_0, \tilde y_t(\beta), y_b(\beta)) =\nabla_{\gamma_i}  
\tilde v(\gamma_i; \beta_0(\gamma_i))$ where 
$\tilde v \equiv -m^2/\lambda$ has the  arguments shown
and $\nabla_{\gamma_i}$ is the gradient in coordinate space  $\{\ln\gamma_i\}$.

The important result is  that $p(D)$ contains a suppression factor $1/\tilde\Delta_q$
where we denoted
\medskip
\bea\label{p001}
&&\tilde \Delta_q(\gamma_i)
\equiv
\big[(f_2)_{\beta}^\prime\big]_{\beta=\beta_0(\gamma_i)}
\,\big\vert \nabla_{\gamma_i}
 \ln \tilde v (\gamma_i; \beta_0(\gamma_i))\big\vert
=
\Delta_q\,\,
\nonumber\\[7pt]
&&\qquad
{\Rightarrow}\quad
p(D)\sim \int dS_{\gamma} \,\,\frac{1}{\Delta_q}\,\,\cL\,\,\times (priors),
\eea
with
\bea\label{deltaq}
\Delta_q=\Big(\sum_{j=1}^5\,\Delta_{\gamma_j}^2\Big)^{1/2},\qquad  
\Delta_{\gamma_j} =\frac{\partial
 \ln \tilde v (\gamma_k;\beta_0(\gamma_k))}{\partial \ln \gamma_j};\qquad 
\gamma_j\equiv m_0, m_{1/2}, \mu_0, A_0, B_0.
\eea

Note that $\tilde\Delta_q(\gamma_i)$ contains a derivative of $\tilde v\sim f_1$ evaluated at
 $\beta=\beta_0(\gamma_i)$, so it
encodes the effects of variations about 
the ground state  of  both minimum conditions (\ref{min}), see the two
Dirac $\delta$'s in (\ref{twodeltas}).  A good stability of these conditions 
under such (quantum) variations requires small $\tilde\Delta_q$.
Interestingly we also notice that $\tilde\Delta_q(\gamma_i)=\Delta_q$ so $\Delta_q$ is
preferred by the calculation of the Bayesian evidence $p(D)$.
The points $\{\gamma_i\}$ of smaller $\Delta_q$, give larger
contribution to $p(D)$, but this also depends on $\cL$ or priors.
We can say that $1/\Delta_q$ is an extra effective prior, emerging when 
marginalizing over parameters, subject to the theoretical constraints. 
 With $p(D)\!\sim\! 1/\Delta_q$, 
points of large $\Delta_q$  pay the fine-tuning cost and so
have a small impact  on  $p(D)$. The latter is then used to decide which of 
two models is more probable.

For further illustration, assume log priors for 
Yukawa couplings  $p(y_{t(b)})=1/y_{t(b)}$ and for SUSY parameters $\{\gamma_i\}$,
using $p(\gamma_1...\gamma_5)=p(\gamma_1)...p(\gamma_5)$ and 
with $p(\gamma_i)=1/\gamma_i$. Then 
\medskip
\bea\label{p1}
p(D) &=& \frac{N}{ 2 v_0 m_Z^0 m_b^0 m_t^0}\,
\int_{\cM}
 d S_{\gamma}\,\,\,\frac{1}{\Delta_q(\gamma_i)}\,
\cL\big( D\vert \gamma_i;\beta,\, v_0,\, \tilde y_t(\beta),\,
 \tilde y_b(\beta)\big)\Big\vert_{\beta=\beta_0(\gamma_i)}.
\eea

\medskip\noindent
To conclude, $1/\Delta_q$ is an extra {\it effective} prior 
$\tilde p_{eff}(\gamma_i)$ of the model and $\Delta_q$ emerges as a measure 
of fine tuning.  In the general case $\tilde p_{eff}(\gamma_i)$ can be read from 
 (\ref{pofD}), (\ref{p001}) and the link between $\Delta_q$ and the Bayesian evidence $p(D)$ 
is clear. Numerical studies of $p(D)$  
or $\cL(D\vert\gamma_i)$
should then include such effect due to the two theoretical constraints.  {
To our knowledge this effect was so far overlooked in such studies.}

Note that  
$\Delta_q$ that emerges in eqs.(\ref{p001}), (\ref{p1}),
 does not contain partial derivatives wrt Yukawa couplings.
This is because  {these} are nuisance (SM-like) parameters that were integrated out,
so are included as a global effect. 
Also, such parameters are not part of the {\it new} ones ($\gamma_i$)  
that SUSY  introduces, so it is no  
surprise that they are not explicitly manifest in $p(D)$ or in
the  denominator under integrals (\ref{p001}), (\ref{p1}).

\begin{figure}[t!]\begin{center}
\includegraphics[width=14.cm,height=4.8cm]{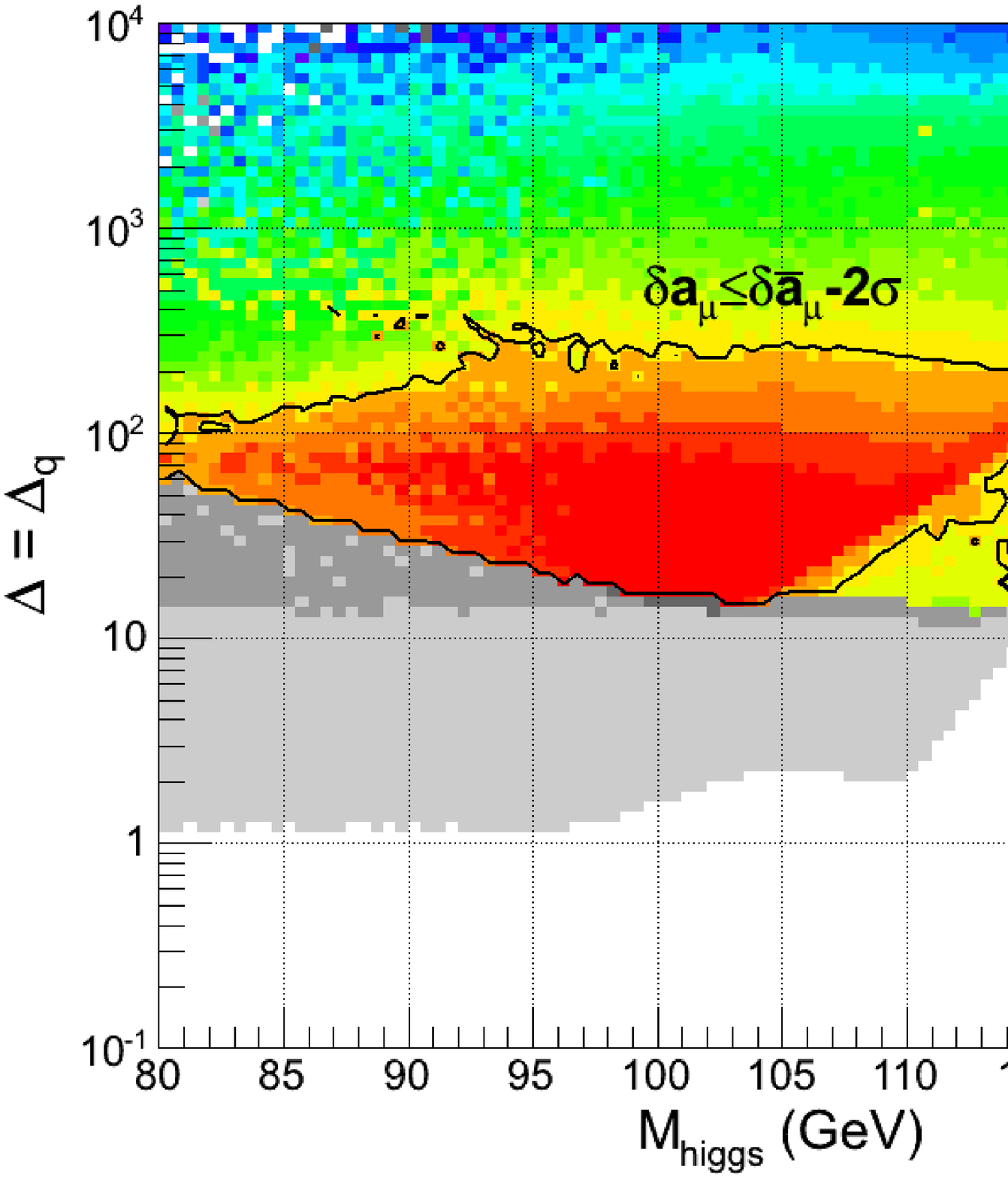}
\includegraphics[width=14.cm,height=4.8cm]{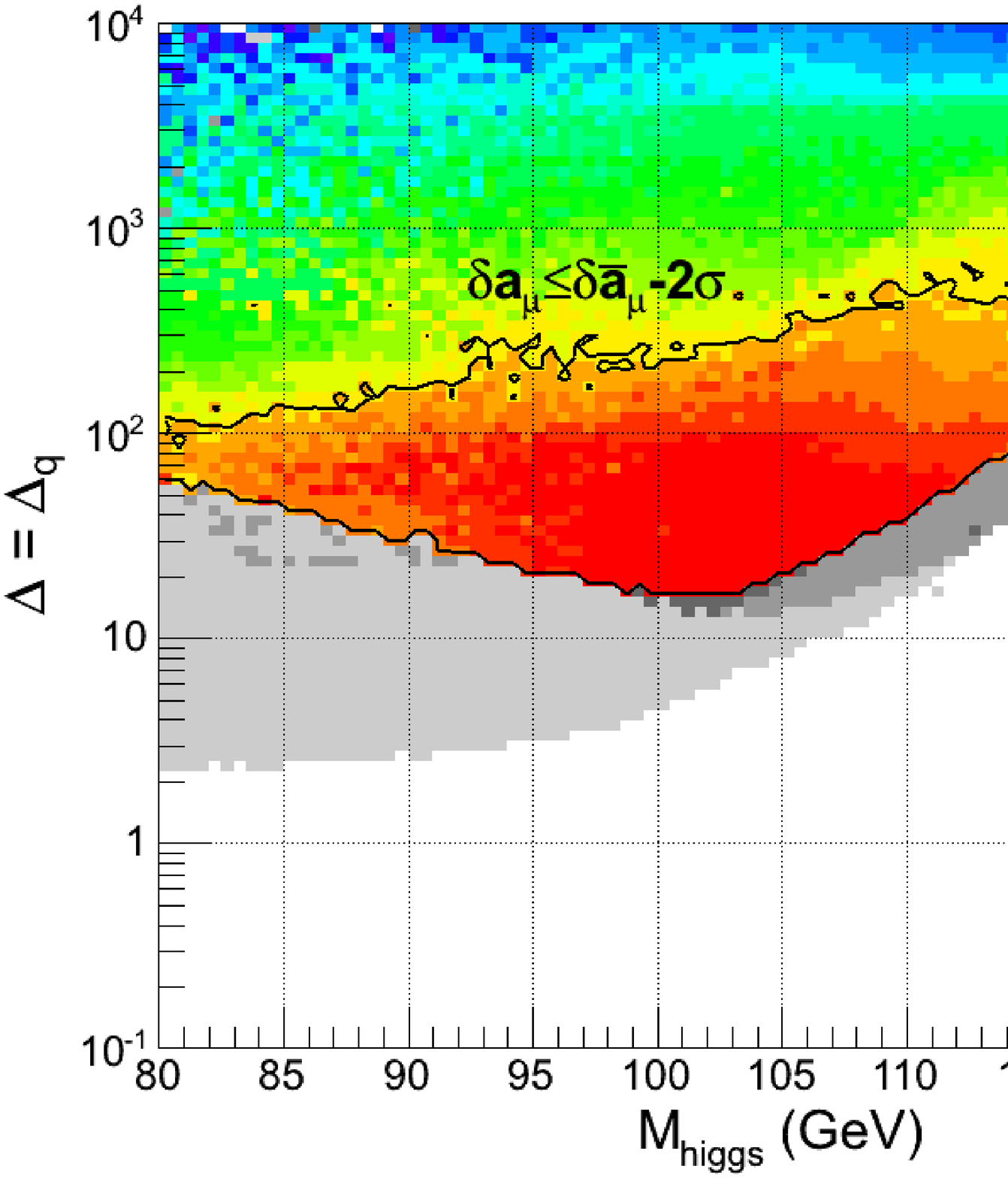}
\includegraphics[width=14.cm,height=4.8cm]{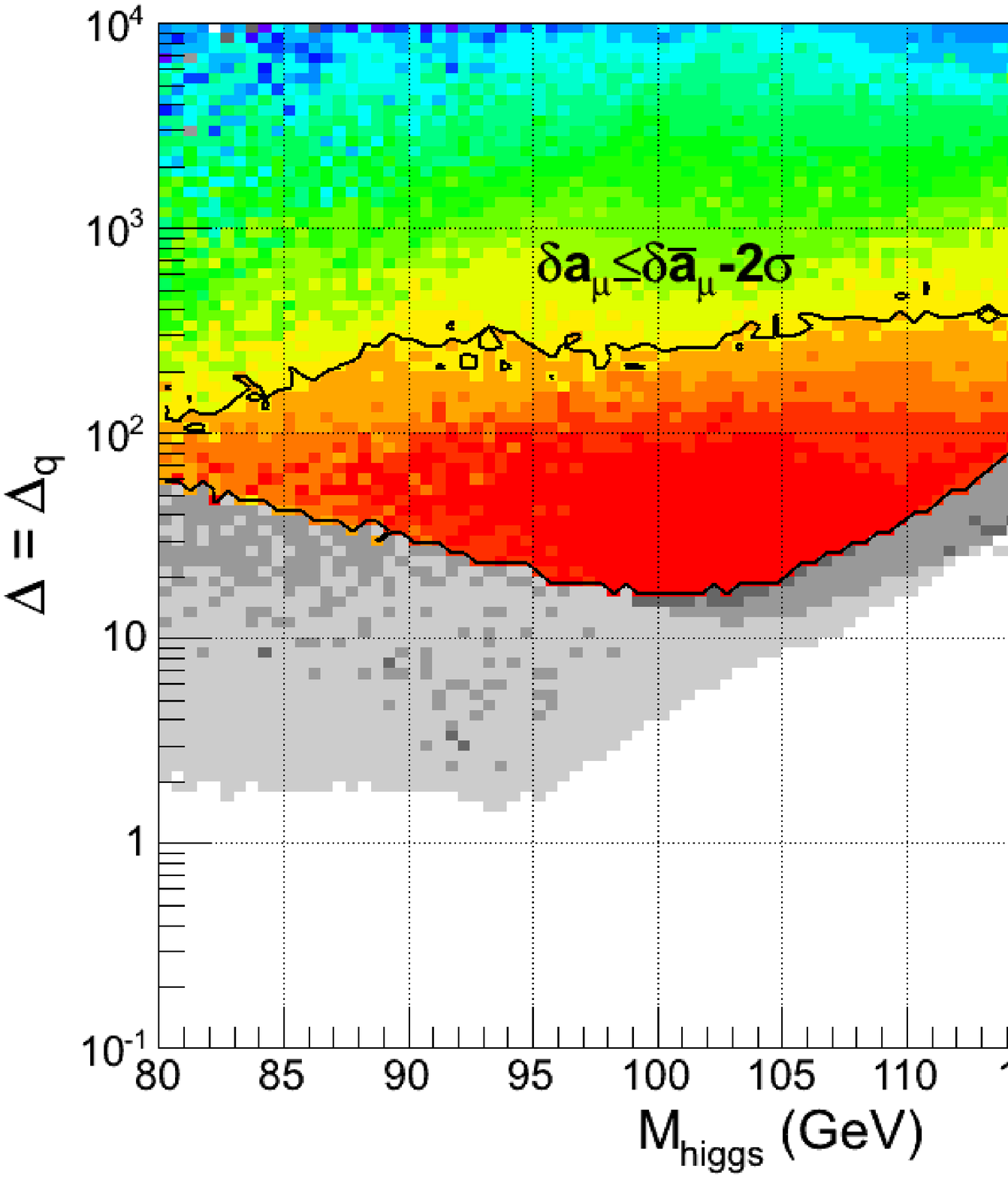}
\includegraphics[width=14.cm,height=4.8cm]{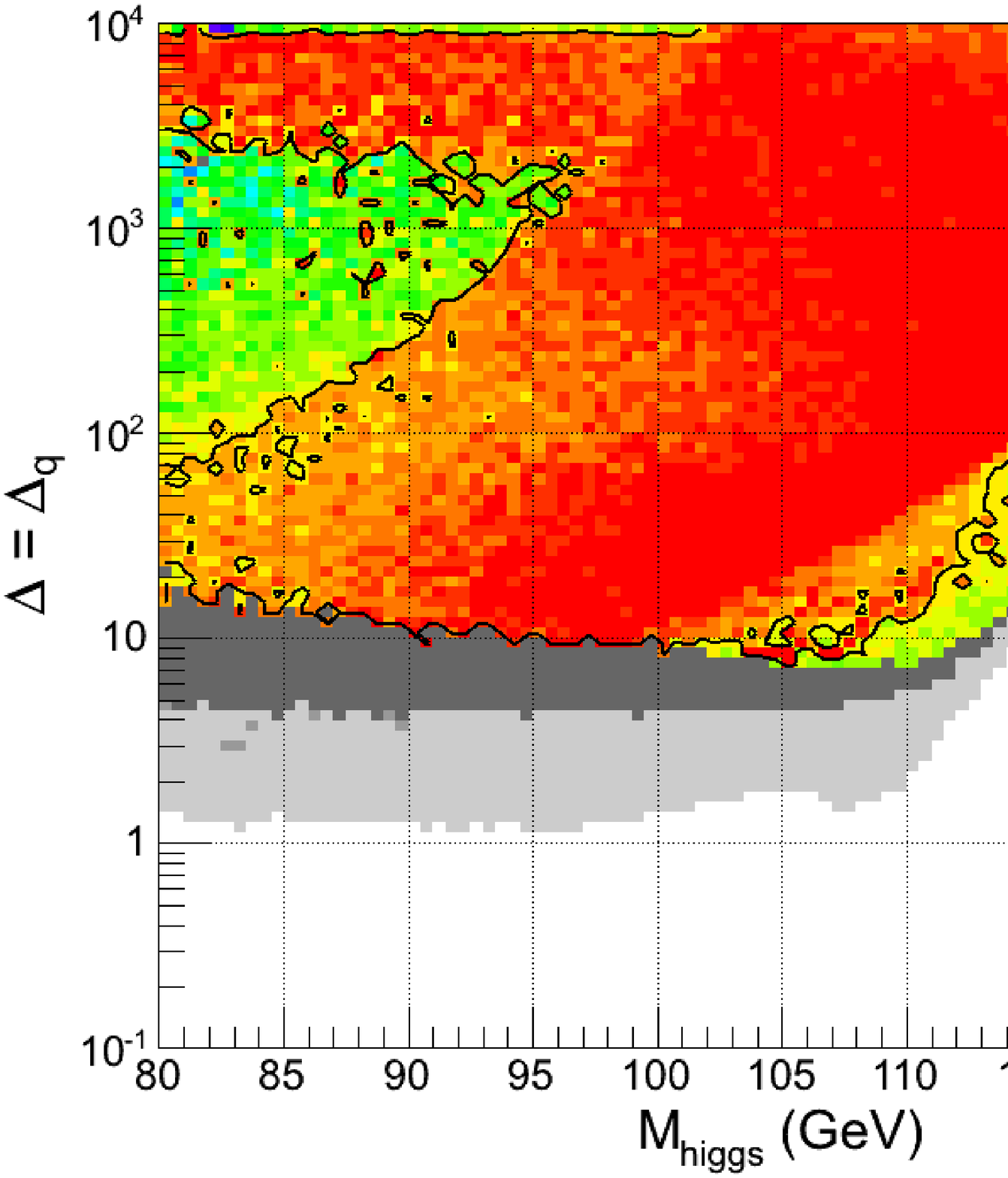}
\end{center}
\renewcommand{\baselinestretch}{0.9}
\vspace{-0.2cm}
\caption*{\small Figures 1 to 4: $\Delta_q$, $\Delta_{max}$ 
versus $M_{higgs}$; lightest grey (0)  area:  excluded by SUSY mass bounds; 
darker grey (1): excluded by $b\!\rightarrow \! s \gamma$, 
$B\!\rightarrow\! \mu^+\mu^-$\!, $\delta\rho$; dark grey (2): excluded 
by condition  $\delta a_\mu \geq 0$.
Coloured area: allowed by data and $\delta a_\mu\leq (25.5 + 2\!\times\! 8) 10^{-10}$;
$\delta a_\mu^{max}$ is shown colour encoded.
Area outside contour: $\delta a_\mu^{max}\!\leq\! (25.5 - 2\!\times\! 8) 10^{-10}$
($2\sigma$). Red area (inside): {\it largest} 
$\delta a_\mu$ is within $2\sigma$ of $\delta a_\mu^{exp}$.
}
\label{figure14}\end{figure}

These above results 
 bring  {technical}
support to a physical meaning of the fine tuning. 
They show that it is desirable to have a smaller
$\Delta$, as also expected from  physical considerations.
Again, one should remember that this may not
always be the region from where $p(D)$ receives the largest 
contributions, as this depends also on the priors, the integral(s) 
or their measure. 
Note also that
changing the priors of the nuisance parameters or 
the measure 
can give  different values for  {Bayesian} $p(D)$
although with enough data $D$ one expects this dependence to become weaker. 
With this technical motivation for the fine tuning measures and their 
relation to $p(D)$, $\cL(D\vert\gamma_i)$
  below we study the values of $\Delta_q$, $\Delta_{max}$
for many SUSY models.

\section{Numerical results for $\Delta$ in generic supersymmetric models.}
\label{num}

We present our numerical results for $\Delta_q$ and $\Delta_{max}$
in a comparative analysis for generic models used for SUSY searches at the LHC. 
We scan the entire parameter space $\{\gamma_i\}$ of the models, consistent with the 
theoretical constraints, using a two-loop leading-log analysis.  
 $\Delta_q$ and $\Delta_{max}$ are presented as functions of physical 
scales  (mass of higgs, stop, gluino, SUSY 
scale  $m_{susy}=(m_{\tilde t_1} m_{\tilde t_2})^{1/2}$) 
with constraints (muon magnetic momentum $\delta a_\mu$,  dark matter abundance $\Omega h^2$). 
The models considered are:

\medskip\noindent
$\bullet$ the CMSSM model, 
of parameters $\gamma_j \equiv \{m_0, m_{1/2}, \mu_0, A_0, B_0\}$.
Then $\Delta_q$ is that shown in (\ref{tun}) and (\ref{deltaq}), 
evaluated at the two-loop leading log level.
See \cite{Cassel:2010px} for a recent study, 
whose results were  recovered by this work.

\medskip\noindent
$\bullet$  the NUHM1 model:  this is a CMSSM-like model but with  Higgs
 masses in the ultraviolet (uv) different from $m_0$, 
$m_{h_1}^{uv}=m_{h_2}^{uv}\not= m_0$, with parameters
$\gamma_j\equiv \{m_0, m_{1/2}, \mu_0, A_0, B_0, m_{h_1}^{uv} \}$.
Then $\Delta_q$ is as in (\ref{deltaq}) with summation over this set.

\medskip\noindent
$\bullet$ 
the NUHM2 model: this is a CMSSM-like model with non-universal Higgs mass, 
$m_{h_1}^{uv}\!\not=\!m_{h_2}^{uv}\!\not=\! m_0$,  
with independent parameters  $\gamma_j\equiv 
\{m_0, m_{1/2}, \mu_0, A_0, B_0, m_{h_1}^{uv},m_{h_2}^{uv}\}$.
Then  $\Delta_q$ is that of (\ref{deltaq})  with summation over this set.

\medskip\noindent
$\bullet$ the NUGM model:
 this is a CMSSM-like model with non-universal gaugino masses 
$m_{\lambda_i}$, $i=1,2,3$,  with 
$\gamma_j=\{m_0, \mu_0, A_0, B_0, m_{\lambda_1}$, $m_{\lambda_2}, m_{\lambda_3}\}$.
Then $\Delta_q$ is given by (\ref{deltaq})  with the sum over this set.

\medskip\noindent
$\bullet$ the NUGMd model: this is a special case of 
NUGM-like  model with a relation among 
the gaugino masses $m_{\lambda_i}$, $i=1,2,3$,  of the type 
$m_{\lambda_i}=\eta_i\, m_{1/2}$,
where $\eta_{1,2,3}$ take only {\it discrete}, fixed values. Such relations can exist
due to some GUT symmetries, like SU(5), SO(10), etc. 
The particular relation we consider is a benchmark point 
of \cite{Horton:2009ed}  with 
$m_{\lambda_3}=(1/3) m_{1/2}$, 
$m_{\lambda_1}=(-5/3) m_{1/2}$, 
$m_{\lambda_2}=m_{1/2}$, corresponding to a particular GUT (SU(5)) model, 
see Table~2 in \cite{Horton:2009ed}.
As a result,  $\Delta_q$ is that of (\ref{deltaq}) with 
$\gamma_j=\{m_0, m_{1/2}, A_0, B_0, \mu_0\}$.

In all models we also evaluate the alternative definition of $\Delta$ given by 
\bea
\Delta_{max}=\max\vert {\Delta_{\gamma}}\vert ,  
\qquad {\textrm{$\gamma$: parameters of  mass dimension 1.}}
\eea
and where the set $\gamma_j$ is listed above for each model.

\begin{figure}[t!]\begin{center}
\includegraphics[width=14.cm,height=4.8cm]{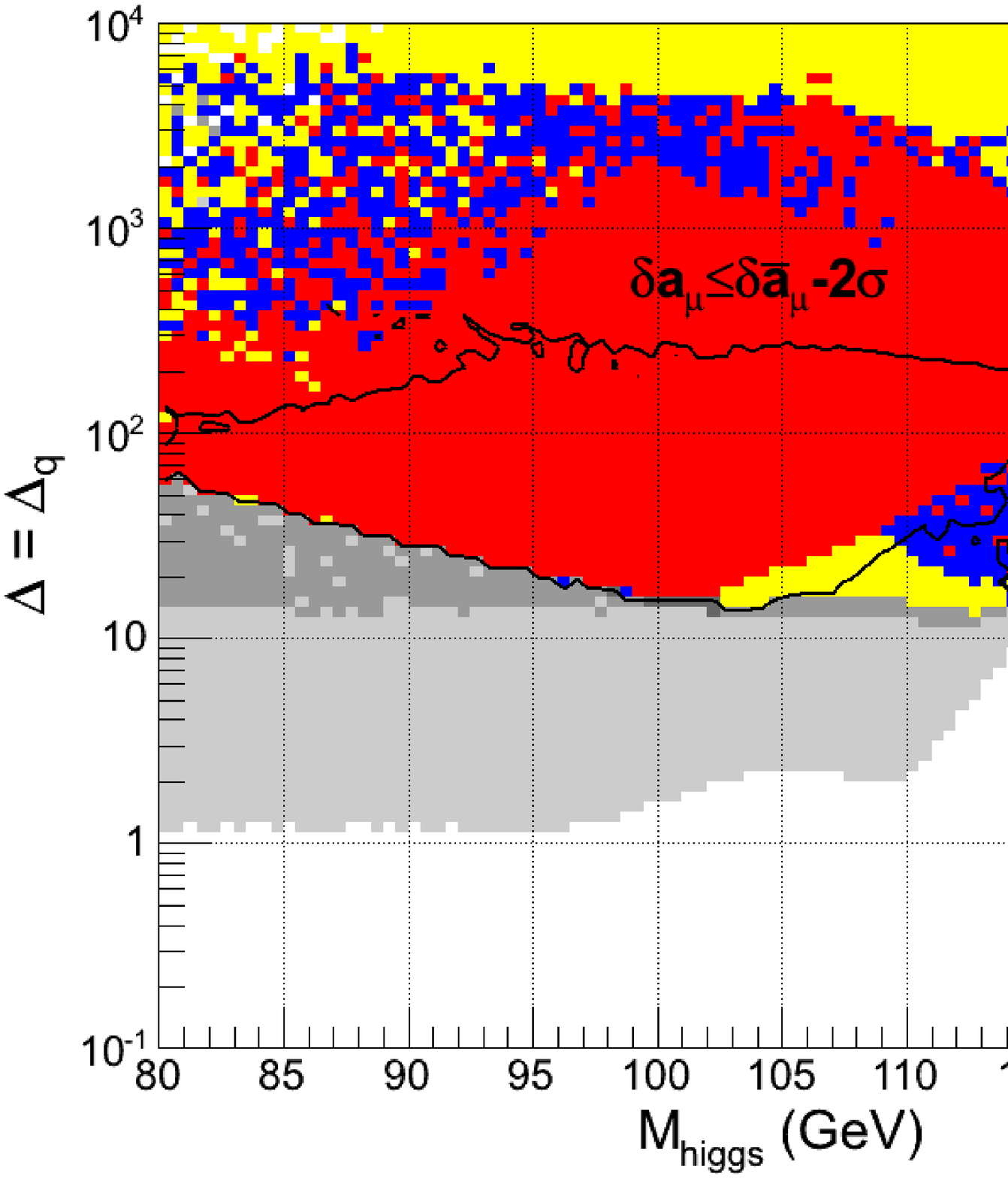}
\includegraphics[width=14.cm,height=4.8cm]{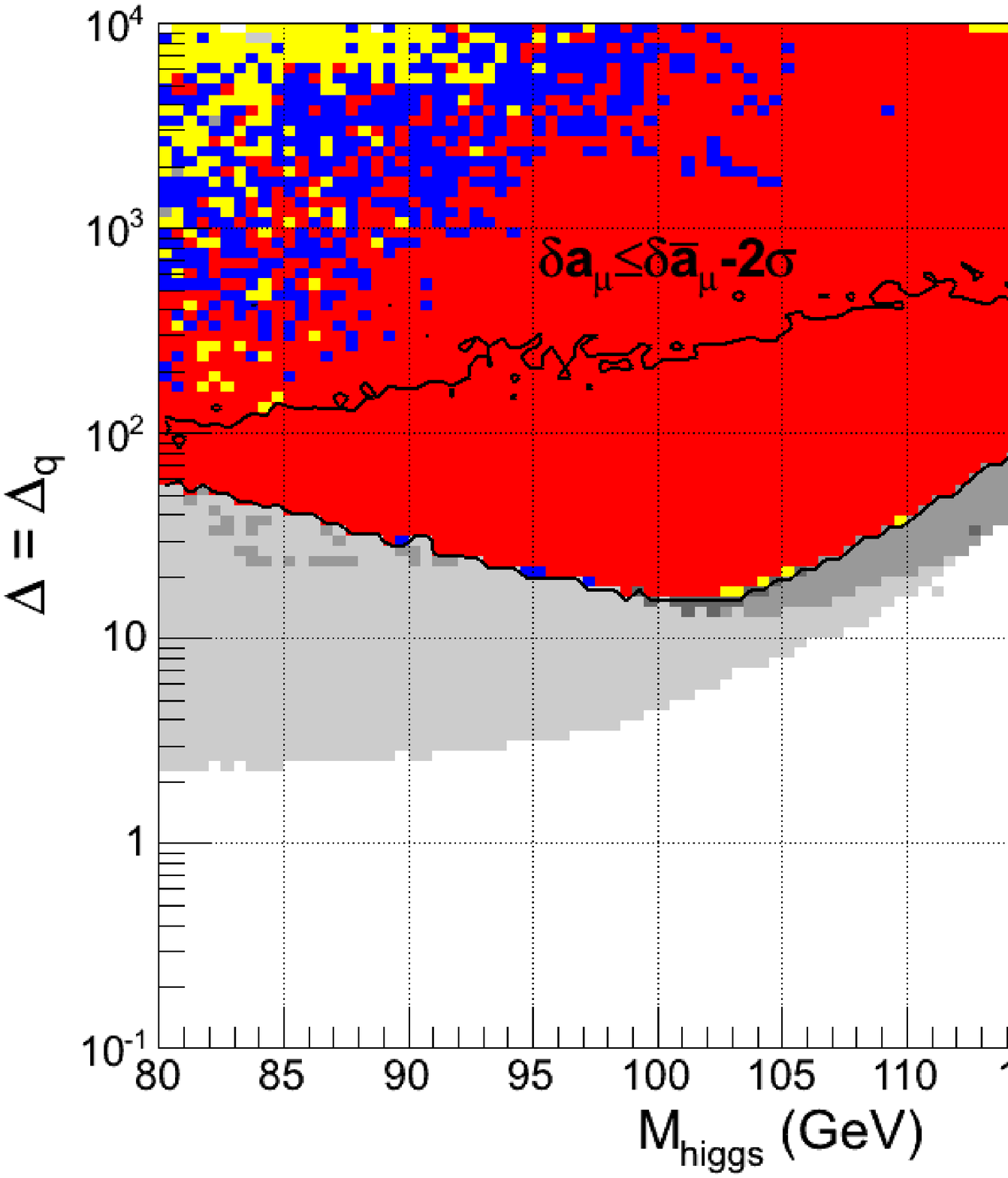}
\includegraphics[width=14.cm,height=4.8cm]{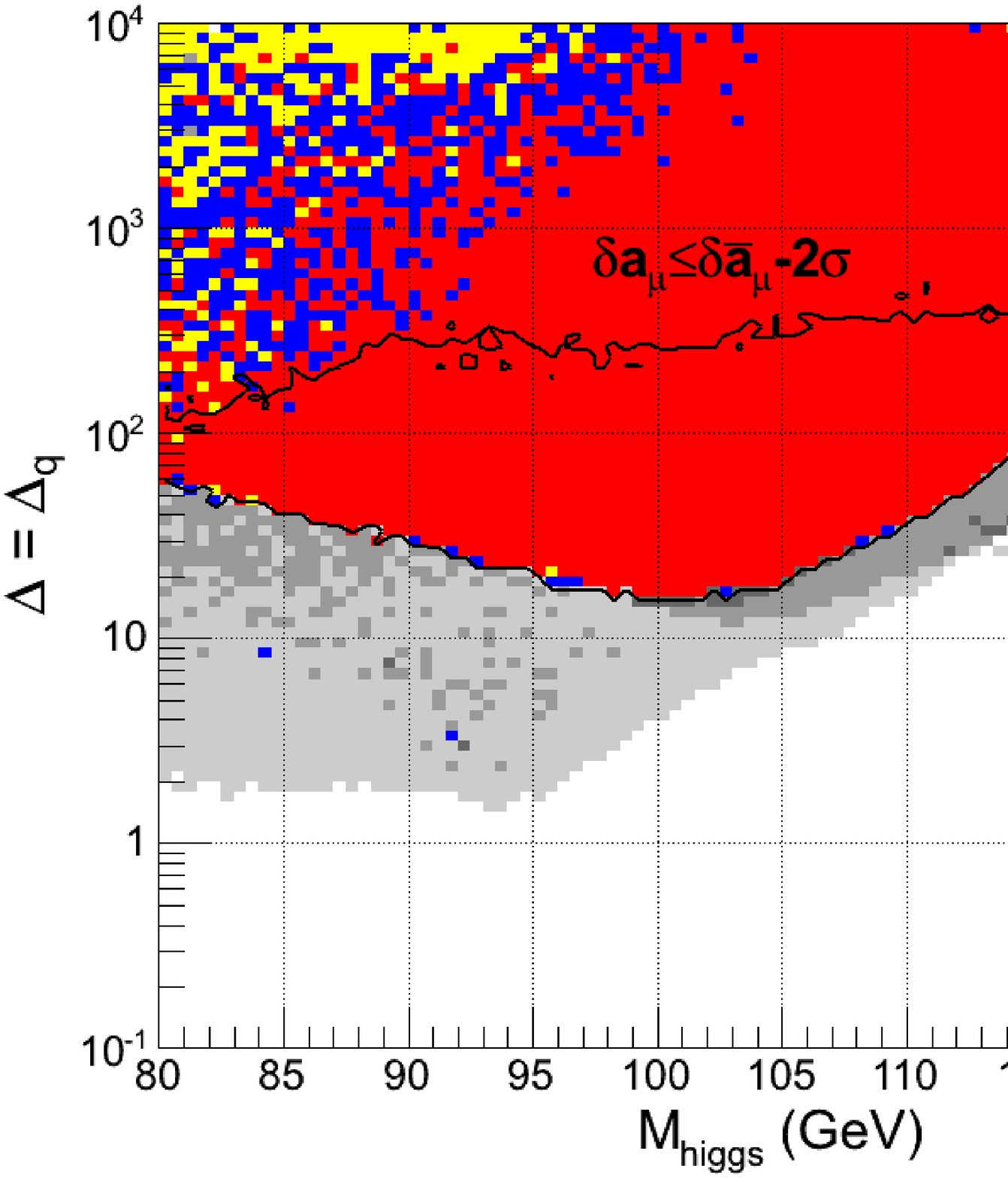}
\includegraphics[width=14.cm,height=4.8cm]{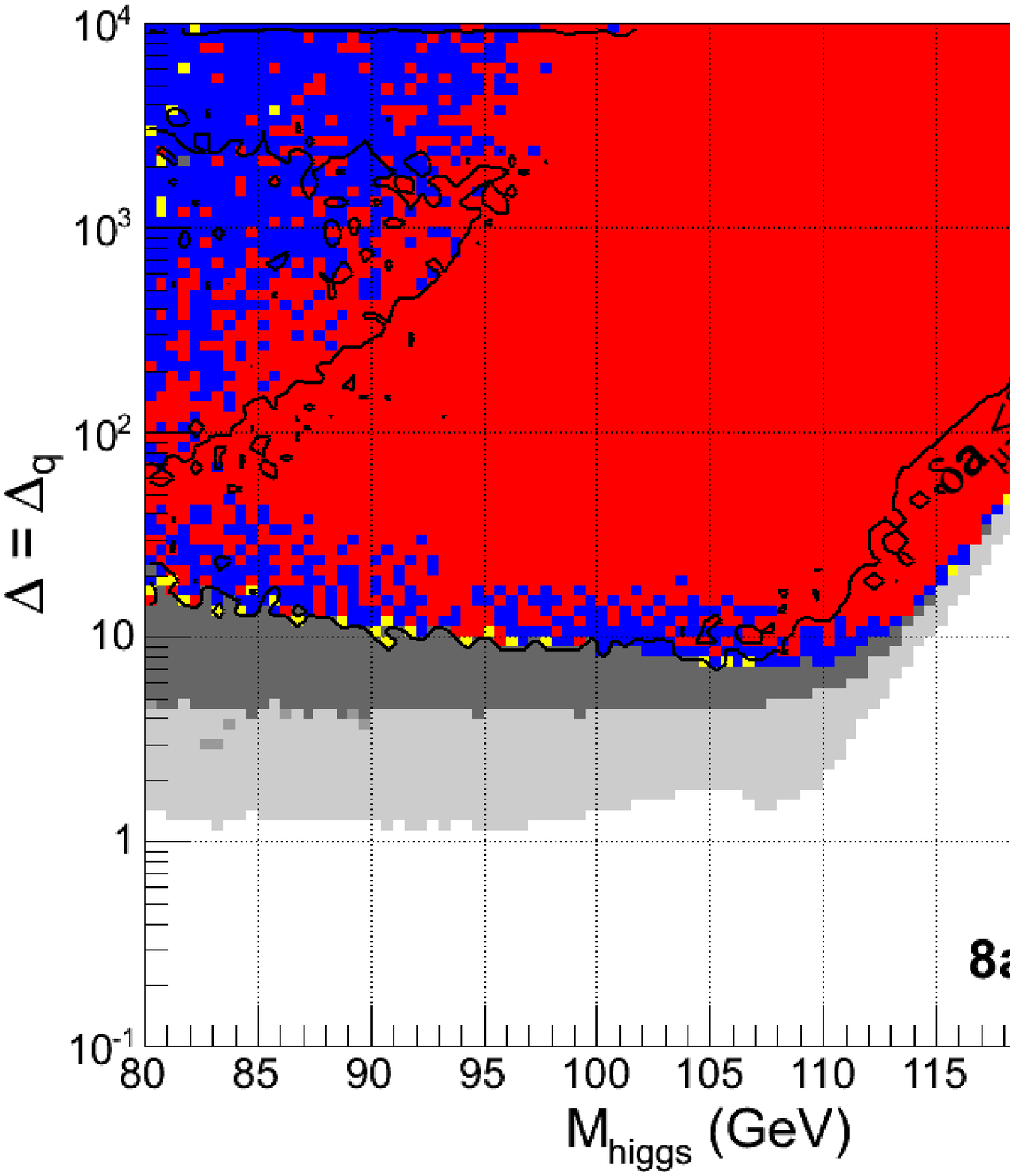}
\end{center}
\renewcommand{\baselinestretch}{0.85}
\caption*{\small Figures 5 to 8:
$\Delta_q$, $\Delta_{max}$ versus $M_{higgs}$:
Various grey areas and $\delta a_\mu$ values: as for Figures~1~to~4.
Colored ares: allowed by data other than $\delta a_\mu$. 
Blue area:  $\Omega h^2\leq 0.1099 - 3\times 0.0062$. Red area
 $0.1099-3\times 0.0062\leq \Omega h^2\leq 0.1099+3\times 0.0062$ ($3\sigma$ saturation).
Yellow: $\Omega h^2\geq  0.1099+3\times 0.0062$.}
\label{figure58}\end{figure}

Before presenting our results let us describe the method used. 
The scan over the full phase space of each model was done using
Pythia~8 \cite{Pythia8} random number generator.
 The public code micrOMEGAs~2.4.5 \cite{Belanger:2001fz} and
 SoftSusy~3.2.4 code \cite{Allanach:2001kg}  were then used, with the latter 
adapted to compute for all models the fine tuning  of the electroweak scale at 
the  two-loop leading log level (instead of its default, one-loop calculation).  
This includes two-loop tadpoles to the two electroweak minimum 
conditions. The data output was then filtered by the experimental
constraints. The run time to generate the phase space points of the five models
was  about 15000 one-day jobs on the CERN computing service, and each plot 
was generated from $\approx 4\times 10^7$ points in a random scan of the 
parameter  space (for alternative and recent data analysis see
\cite{Buchmueller:2011sw}).
 
\bigskip
\begin{table}[th]
\begin{center}
\begin{tabular}{|c|c|} \hline
Experimental constraints & Values used \\ \hline
SUSY particle masses & Routine in micrOmegas 2.4.5,``MSSM/masslim.c" \\
Muon magnetic moment & 
$\delta a_\mu=(25.5 \pm 2\times 8 ) \times 10^{-10}$  at $2 \sigma$ \cite{gminus2}. \\
$b\rightarrow s \,\gamma$  process  
& $3.03  < 10^{4} ~ \mbox{Br}(b \to s \gamma) < 4.07$ at $2\sigma$  \cite{bsg}.  \\
$B_s \to \mu^+ \mu^-$  process & 
$\mbox{Br} (B_s \to \mu^+ \mu^-) < 1.08 \times 10^{-8}$ at $2 \sigma$ \cite{bmupmum}.\\
$\rho$-parameter    & \,\,\,\,\,
$\,\,\,\,\,-0.0007< \delta \rho < 0.0033$ at $2 \sigma$   \cite{pdg}. \\
Dark matter relic density   
&   $\Omega h^2=0.1099\pm 3 \times 0.0062$ at $3\sigma$ \cite{wmap}.\\ 
\hline
\end{tabular}
\end{center}
\caption{\small Experimental data constraints. 
$\delta a_\mu$ includes the theoretical error and 
is not imposed on the data, but its values are
shown as a contour plot (at $2\sigma$) or colour encoded from 
which larger deviations can be read ($3\sigma$). 
 {For the other processes in the table, 
only the experimental error is 
considered,  and the details of their theoretical calculation are
  provided by micrOMEGAs 2.4.5 \cite{Belanger:2001fz}, see also
its manual for v.2.4 available at http://lapth.in2p3.fr/micromegas/.
The central values for $m_{top}=173.1$ GeV and 
$\alpha_3(m_Z^0)=0.1184$ \cite{pdg} were  used as inputs in SOFTSUSY.
Note that a combined $1\sigma$ increase of top mass and $1\sigma$ 
decrease of $\alpha_3(m_Z^0)$ can decrease $\Delta_{max}$ by a factor 
as large as 2 (best case scenario), see later.}}
\label{constraints}
\end{table}

Our results\footnote{
 {After this work was completed, an updated bound on $B_s\ra \mu^+\mu^-$ was published 
\cite{Aaij:2012ac}.
We checked that our fine tuning estimates are unchanged,
for a higgs mass in the now preferred region of 122 GeV to 128 GeV.}}
shown in the plots allow the reader to set his 
own constraints on physical scales such as the higgs mass, gluino, stop mass or 
SUSY scale $m_{susy}$, $\delta a_\mu$ or dark matter abundance
and  infer from that the amount of fine tuning. 
 {Note also that  the LEP2 bound on $M_{higgs}$  is never imposed on our figures,
and we let the reader to do this, in the light of future LHC results\footnote{
Note that a flat bound like LEP2 bound on $M_{higgs}$
should be used with care since it applies only to SM.}.}
 This  has the great advantage 
that the impact of  {future}  bounds from LHC on these physical scales 
can very easily be seen on the plots, without the need to re-do the whole analysis.
 Space constraints do not allow us to also present a 
description of the allowed parameter space $\{\gamma_i\}=\{A_0, B_0, etc..\}$
used in these plots, due to complicated correlations among these,
that  can only be presented as more  additional figures, that we postpone
to  a future work. 
Finally, the parameter space ($\{\gamma_i\}$) that we scanned over was: 
$A_0\in[-7, 7]$ TeV, $m_0\in[0.05,5]$ TeV, 
$m_{1/2}\in[0.05,5]$ TeV and also $2\leq \tan\beta\leq 62$. All plots
are marginalized over $\tan\beta$ and $\{\gamma_i\}$.

\begin{figure}[t!]\begin{center}
\includegraphics[width=14.cm,height=4.8cm]{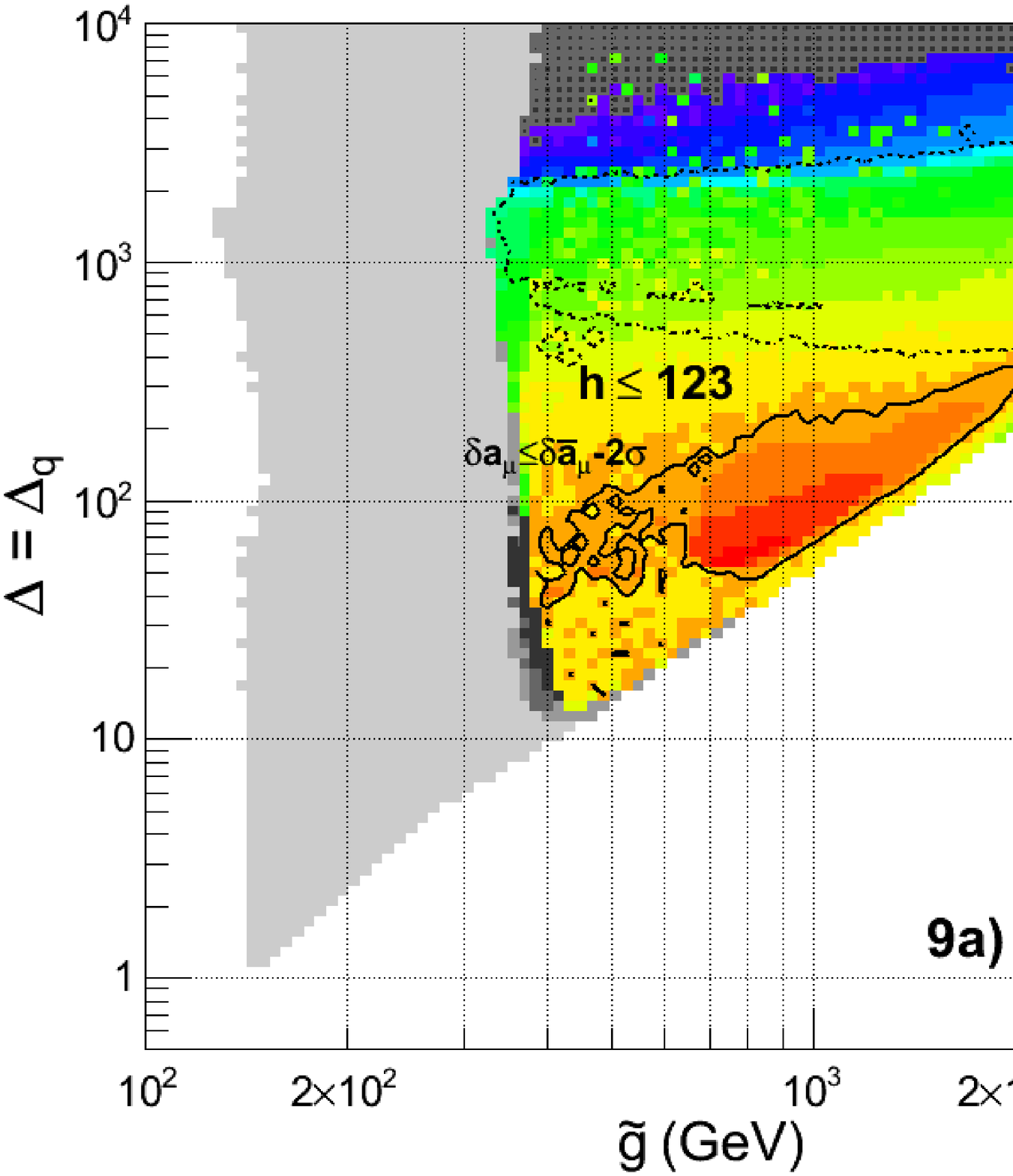}
\includegraphics[width=14.cm,height=4.8cm]{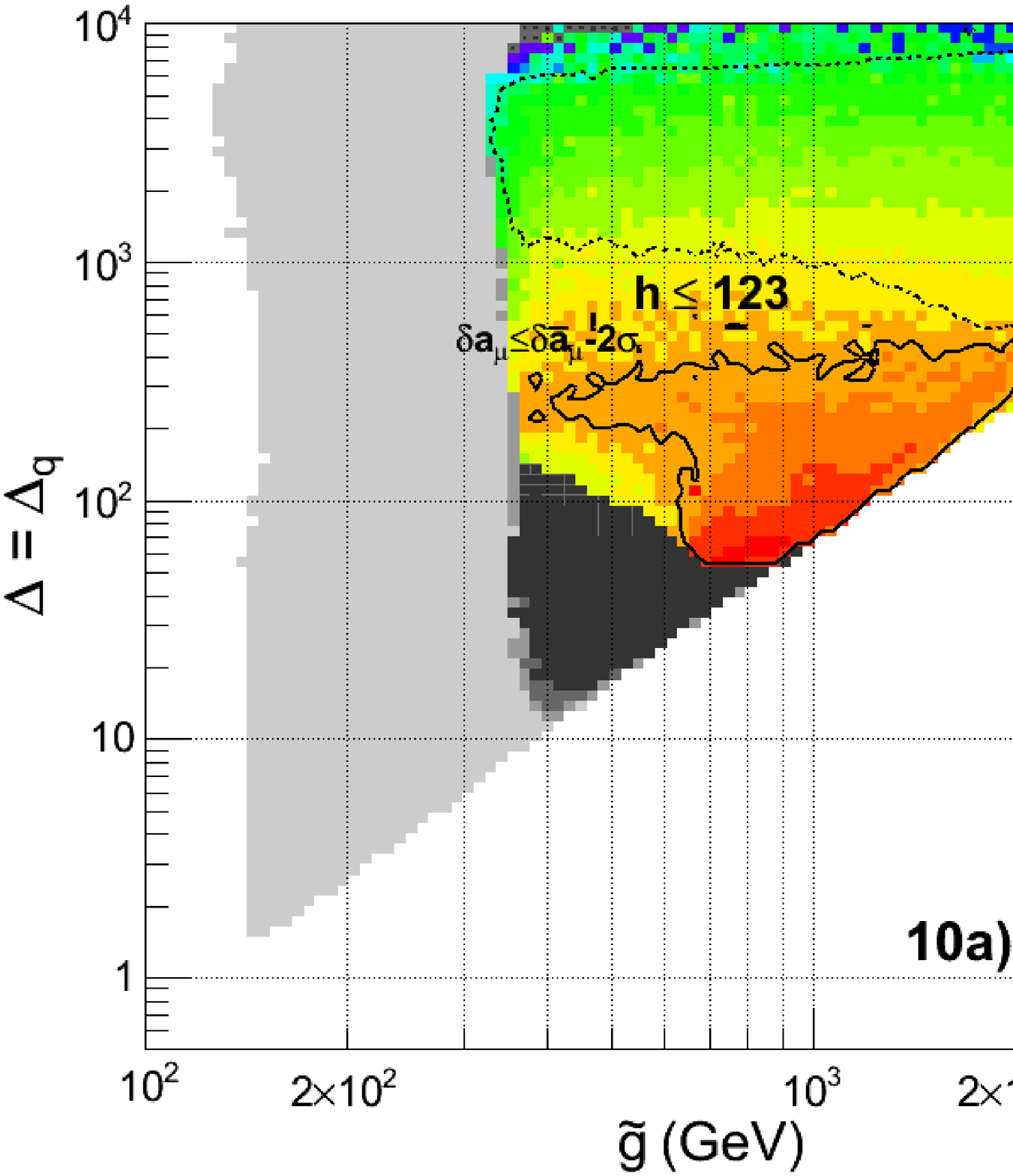}
\includegraphics[width=14.cm,height=4.8cm]{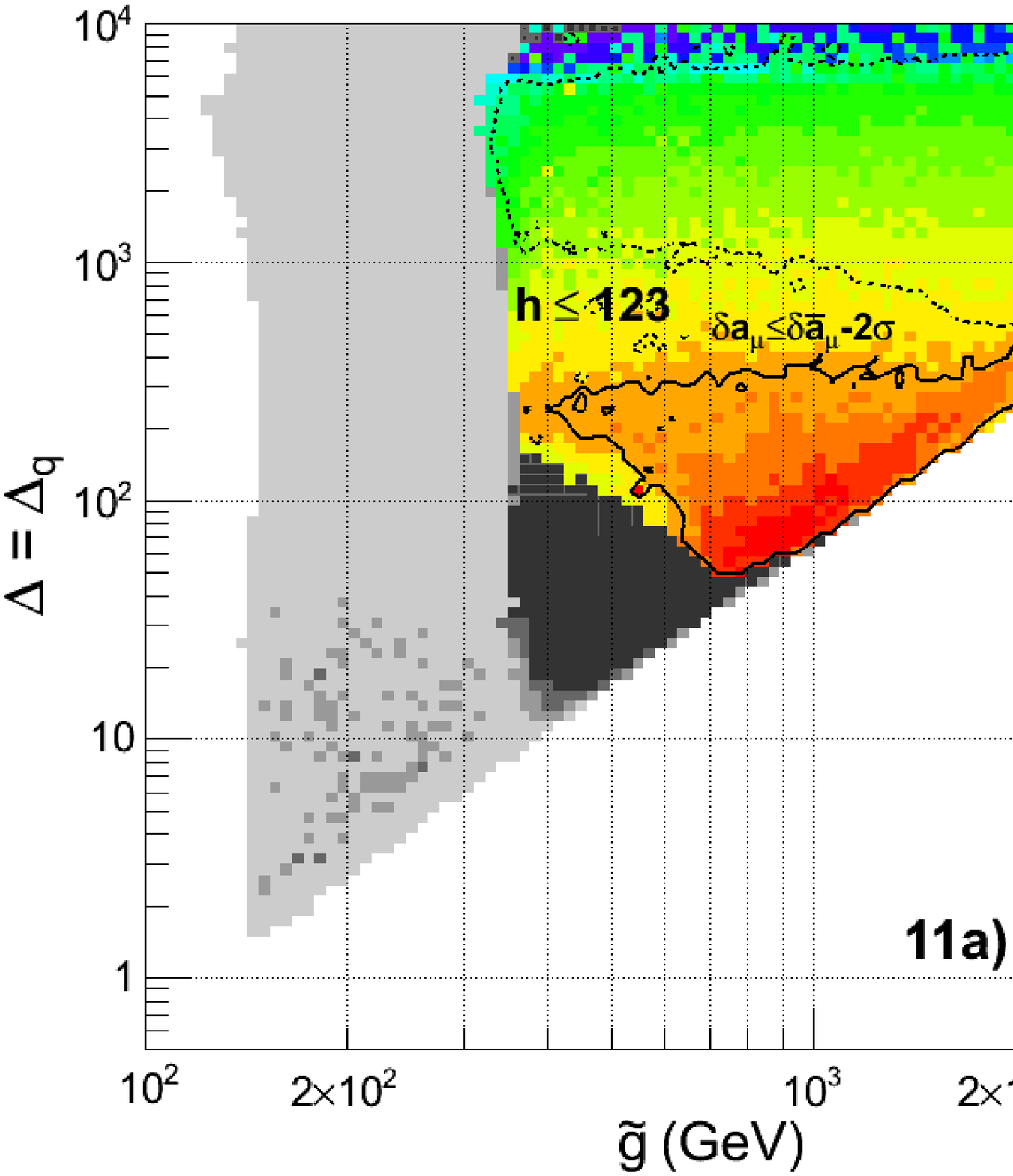}
\includegraphics[width=14.cm,height=4.8cm]{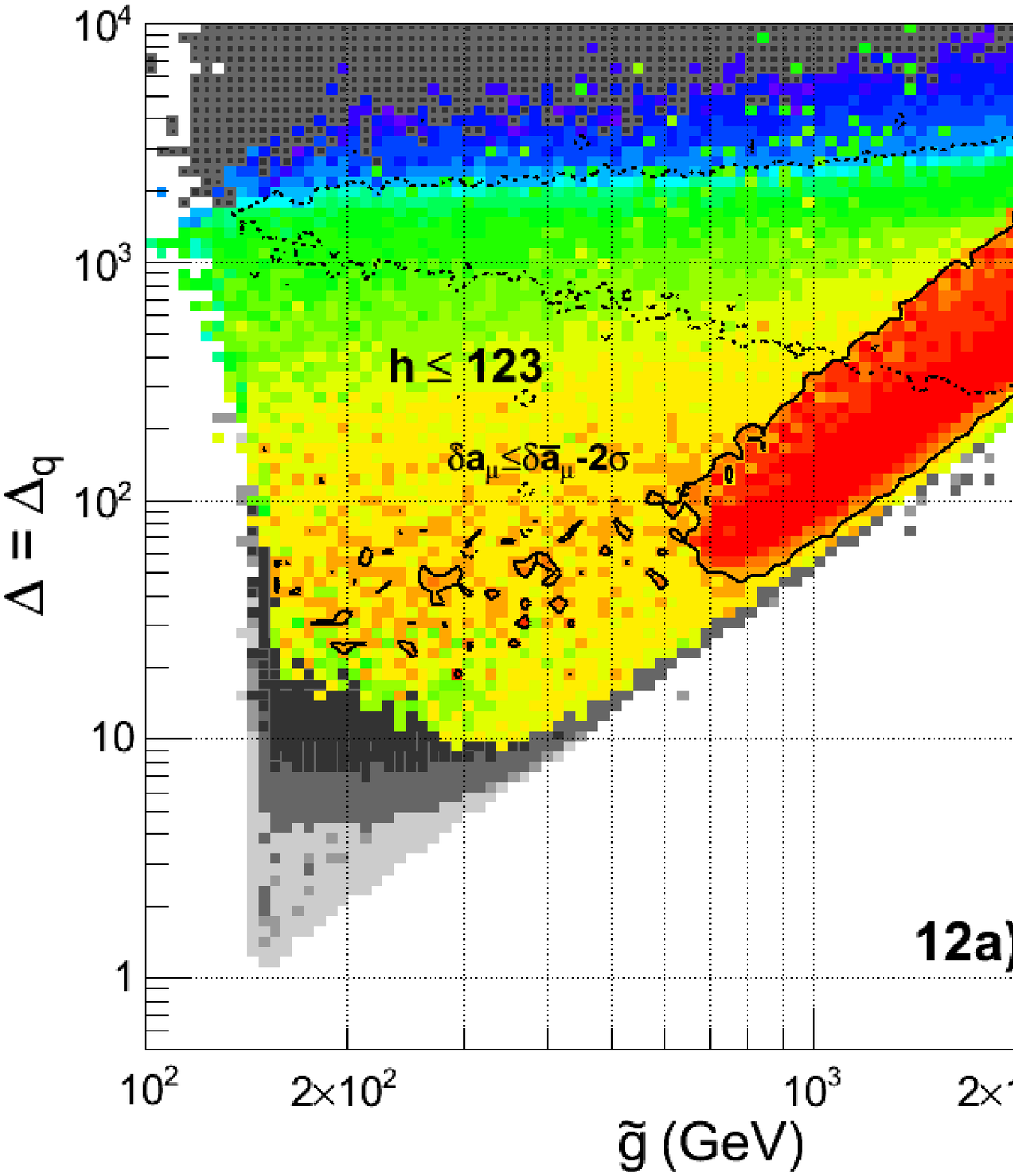}
\end{center}
\renewcommand{\baselinestretch}{0.9}
\caption*{\small Figures 9 to 12:
$\Delta_q$ versus gluino (left) and
SUSY scale (right) for various models; 
$M_{higgs}\!<\!123$ GeV  in area below the dotted line;
similar plots exist for $\Delta_{max}$; various grey
areas as in Figures 1~to~4, forbidden by data. Black area: 
$M_{higgs}<111.4$ or $M_{higgs}\geq 130$ GeV.
Outside the red area:
$\delta a_\mu^{max}\!\leq\! (25-2 \times 8) 10^{-10}$; inside
this area $\delta a_\mu^{max}$ is within $2\sigma$ of $\delta a_\mu^{exp}$.
See also caption of Figs 1-4.}
\label{figure912}\end{figure}

\subsection{$\Delta$ versus $M_{higgs}$ and the values of $\delta a_\mu$.}

In figs.~1 to 4 a), b)   and  17 a), b), 
we show the plots for $\Delta_q$ and $\Delta_{max}$ as functions of the mass of the lightest 
Higgs boson $M_{higgs}$, for all models: CMSSM, NUHM1,  NUHM2, NUGM, NUGMd. 
The impact of $\delta a_{\mu}$ constraint  is also shown with a 
contour line displaying an island of its {\it largest} values, within $2\sigma$
of $\delta a_\mu^{exp}$. For other values ($3\sigma$ deviations, etc), 
the {\it largest} $\delta a_\mu$  is also  
shown  colour encoded, see the scale on the right side of the plots. 
The lightest grey (level 0) areas in these plots are excluded by 
the lower bounds on the  spartners  masses  obtained from negative SUSY searches. 
The darker grey (level 1) areas are excluded by $B_s\!\rightarrow\! \mu^+\mu^-$, 
$b\!\rightarrow\! s \gamma$ and $\delta \rho$ constraints.
 {The dark grey area (level 2) that we also show, visible  only for NUGM model
corresponds to $\delta a_\mu\!<\!0$ and has  $(m_{\lambda_2} \mu)\!<\!0$ and is present at 
$M_{higgs}\leq 115$ GeV. This region is excluded  
by demanding $\delta a_\mu\!>\!0$,  ($(m_{\lambda_2} \mu)\!>\!0$), 
preferred by $\delta a_\mu^{exp}$ data.}

As it is shown in these figures, the LEP2 bound (114.4 GeV) \cite{LEP2} 
on the higgs  mass was  not imposed. Note however that above this 
value both $\Delta$'s are largely independent of the experimental data (ignoring
$\delta a_\mu$)  for all models other than NUHM1, NUHM2; these still have some 
dependence on data (the small grey area). This is interesting and
suggests that the range of values of
 $\Delta_q$, $\Delta_{max}$ can be fixed mainly
by theory and the higgs mass bound, with little or no impact from  other data.

As it is seen from these results, the differences between 
$\Delta_q$ and $\Delta_{max}$ are practically negligible.  
For a given model and a fixed Higgs mass,  there is a relative factor 
 {between 1 and 2} and  which can be safely 
ignored\footnote{ {In general no 
individual $\Delta_{\gamma_i}$ dominates clearly for all higgs masses, see  fig.2 in  
\cite{Cassel:2010px} for the CMSSM.}}.
There is also very similar behavior i.e. various contour lines such as that of
maximal $\delta a_\mu$ are nearly identical
 for both $\Delta_q$, $\Delta_{max}$. This is interesting and shows that one can 
use either definition for fine tuning to obtain a rather similar result.

\begin{figure}[t!]\begin{center}
\includegraphics[width=14.cm,height=4.8cm]{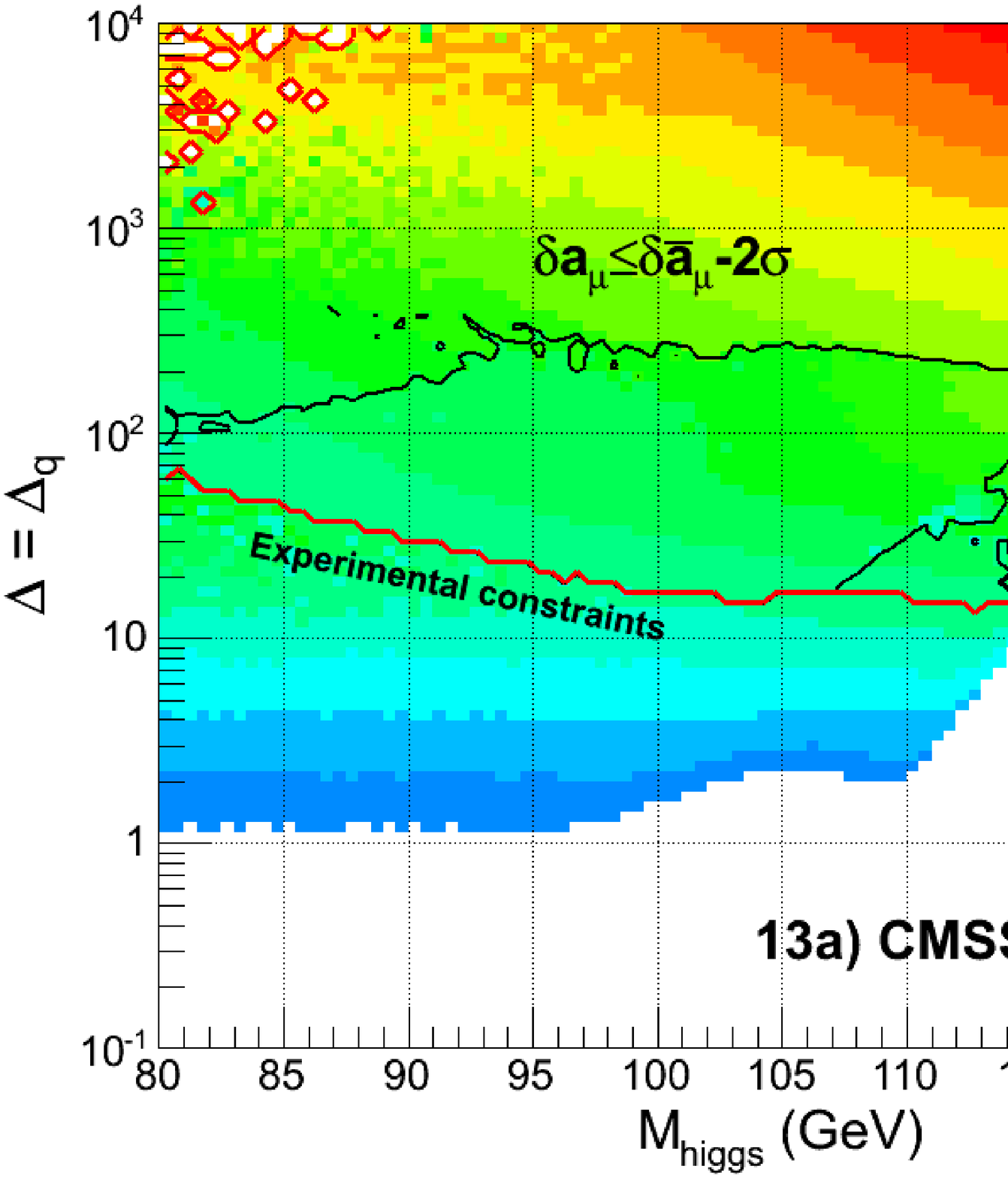}
\includegraphics[width=14.cm,height=4.8cm]{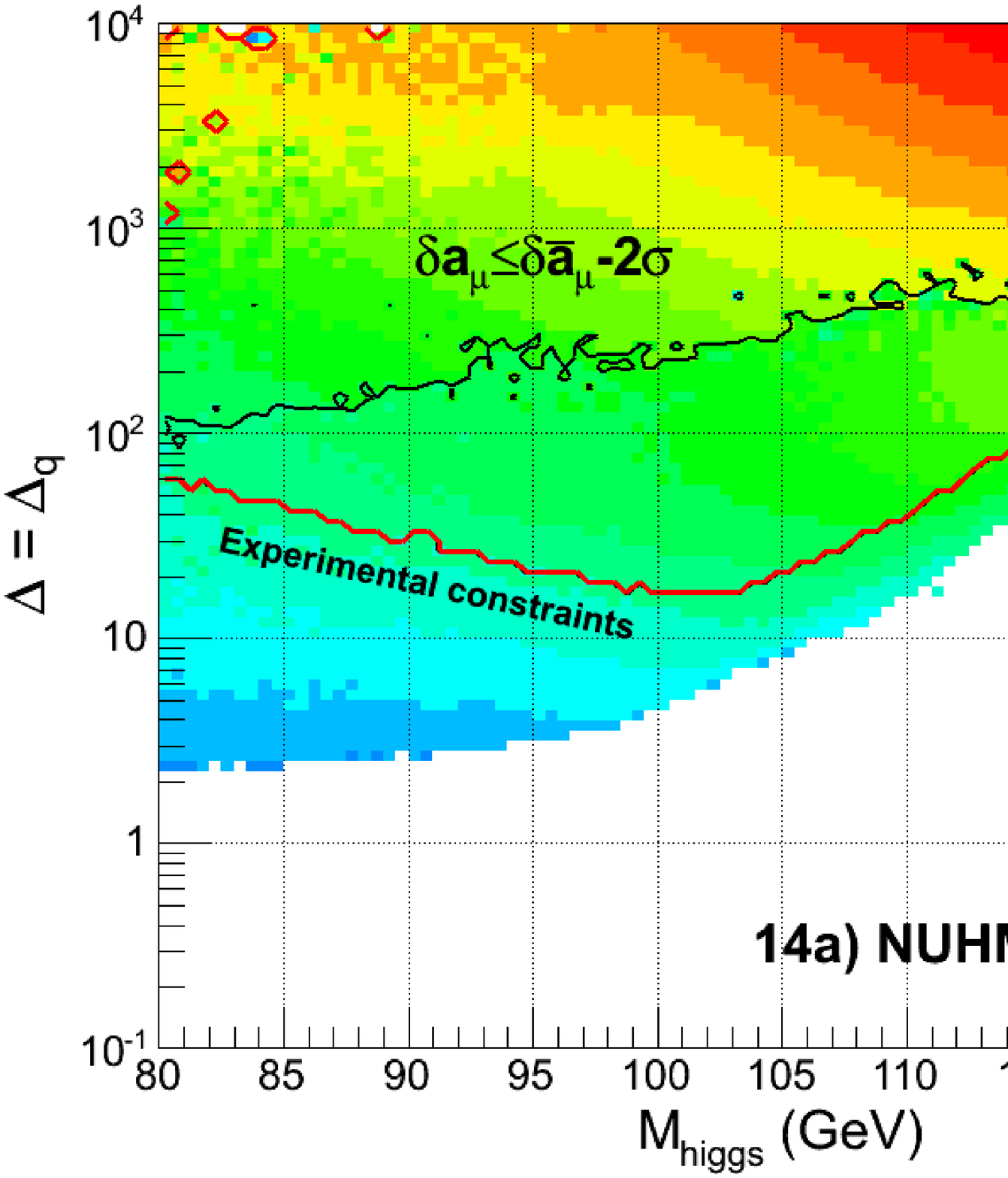}
\includegraphics[width=14.cm,height=4.8cm]{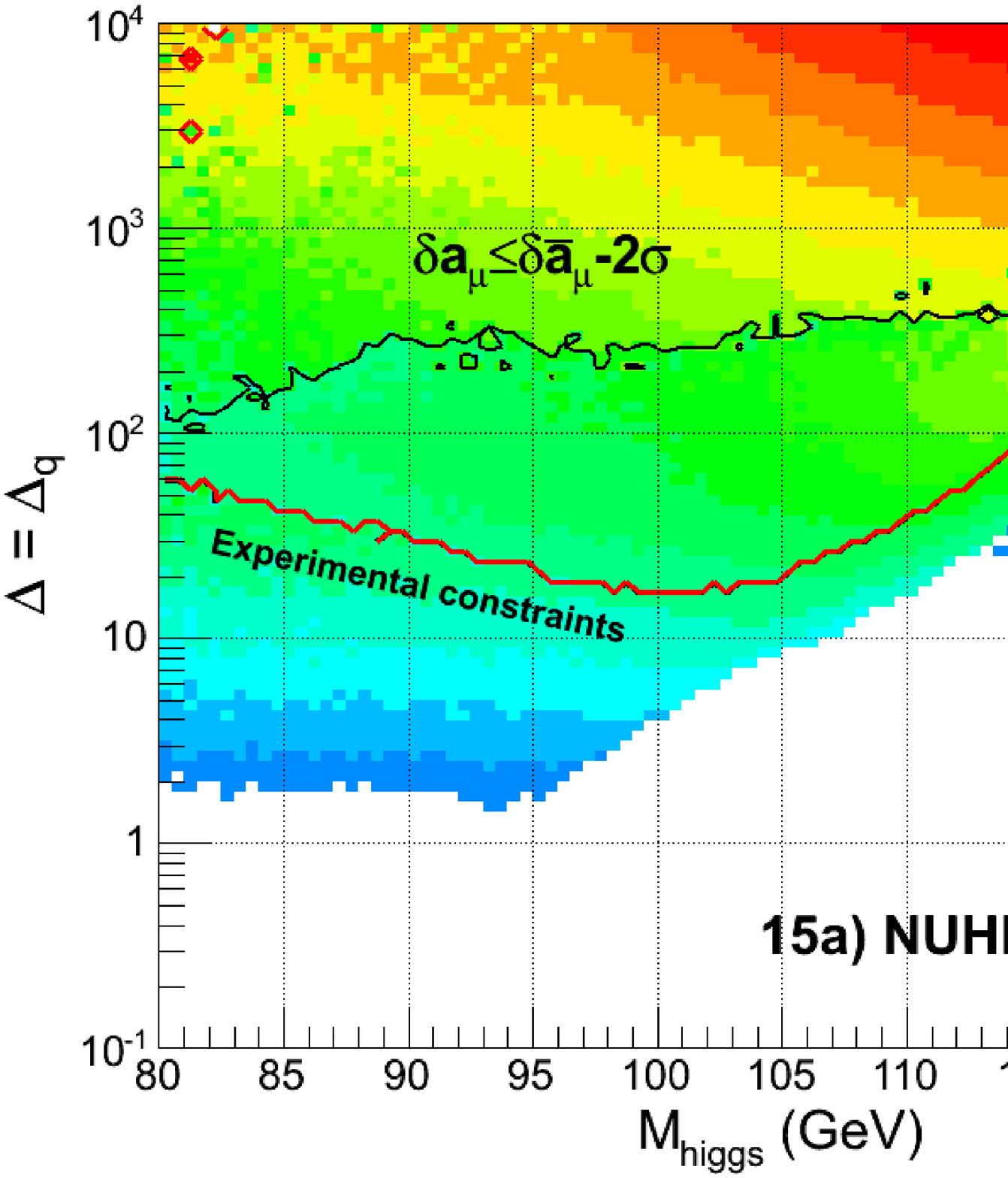}
\includegraphics[width=14.cm,height=4.8cm]{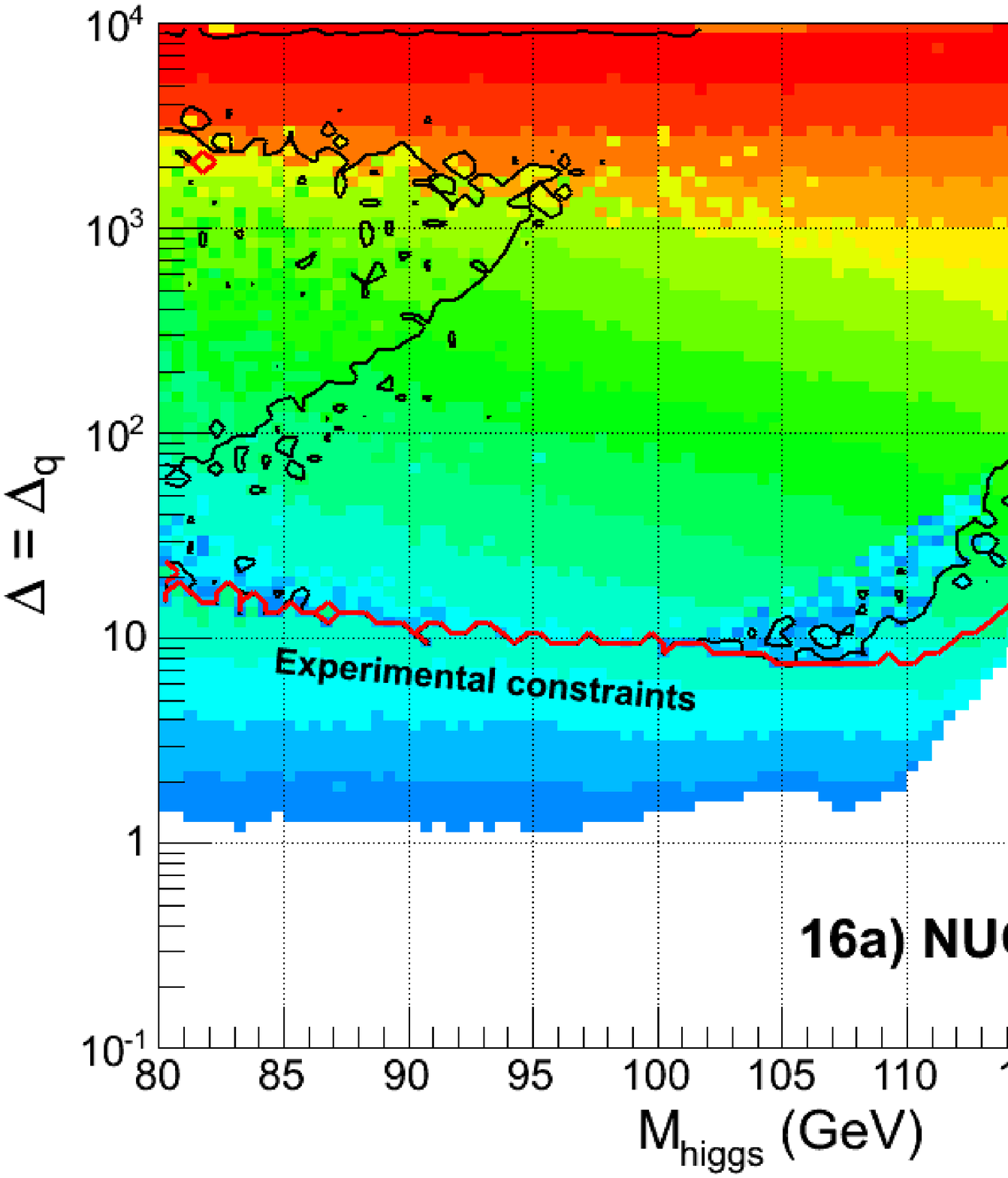}
\end{center}\renewcommand{\baselinestretch}{0.9}
\caption*{\small ‏
Figures 13 to 16: $\Delta_q$ versus $M_{higgs}$ with {\it largest} 
(left plots) and {\it lowest} (right plots) gluino mass; 
For given $\Delta_q, M_{higgs}$ one infers the gluino mass range. 
Area allowed by data  (except $\delta a_{\mu}$) as shown above 
the continuous line,  see also Figures 1-4. 
Values as large as 
 $\delta a_\mu\leq (25- 2\times 8) 10^{-10}$ are outside the closed contour;
inside the contour: {\it largest} $\delta a_\mu$ is within $2\sigma$ of 
$\delta a_\mu^{exp}$ and the gluino mass satisfies it.}
\label{figure1316}\end{figure}

In the CMSSM  the minimal value of 
$\Delta_q$ and $\Delta_{max}$ is situated for higgs mass 
near the LEP2 value  as also discussed in \cite{Cassel:2010px}.
This means that  to respect the LEP2 bound on the Higgs mass
there is no fine tuning cost due to quantum corrections. 
This corrects  common but wrong opposite claims in the literature. 
Further, if one accepted the principle that
$\Delta$ of a model should actually be minimized, 
then one immediately has a CMSSM prediction for
 $M_{higgs}\approx 115\pm 3$ GeV  without using experimental constraints 
(ignoring here $\delta a_\mu$), for details see
 \cite{Cassel:2010px}\footnote{There is a $\pm (2\, {\textrm {to}}\, 3)$ 
GeV theoretical uncertainty from the various public codes
\cite{Allanach:2001kg,PS,SH}.}.
In models other than CMSSM and  after imposing 
the LEP2 bound, the fine tuning is again smallest near this scale.
For $M_{higgs}$ near $115$ GeV, $\Delta_q\approx\Delta_{max}\approx 10$ to $100$,
depending on the model. 
Above this mass value,  both $\Delta_q$, $\Delta_{max}$ grow very fast
($\approx$ exponentially), due to the quantum corrections to the Higgs mass 
(which are logarithmic in $m_{susy}$). As a result, for the currently interesting 
region discussed by CMS and Atlas experiments \cite{atlas2011,CMS2011}, of 
$123\leq m_h \leq 127$ GeV, there is  significant amount of fine tuning required,  
$\Delta_q\approx \Delta_{max}\approx  200$ to $2000$;
 for $M_{higgs}\!=\!125$ GeV,  $\Delta_q\approx \Delta_{max}\approx  500$ to $1000$,
depending on the model.  From these results one could say that 
NUGM is preferable also because it could  more easily comply with  
$\delta a^{exp}_\mu$ ($2\sigma$).
 {Finally, let us mention that a combined $1\sigma$ increase of $m_{top}$
and $1\sigma$ reduction of $\alpha_3(m_Z^0)$ can reduce  (best case scenario)
the fine tuning for a fixed higgs mass by a factor near
$\approx 2$ or so for the CMSSM \cite{Cassel:2010px}, with similar 
effect expected for other models.}

\subsection{$\Delta$ versus $M_{higgs}$ and dark matter relic density.}

Let us now discuss the results of  figures 5-8 a), b)  and 18 a), b).
These present the impact of the dark matter relic density constraint.  
Again, no significant difference between $\Delta_q$ and $\Delta_{max}$ 
is observed for the models considered.
The meaning of light and dark grey areas is the same as in the previous figures. 
In blue we show points
that are consistent with dark matter relic density within  $3\sigma$, i.e. these
points have $\Omega h^2 < 0.1099-3\times 0.0062$. 
The red points saturate the relic density within $ 3\sigma$ deviation from the central
value. Finally, yellow points correspond to a relic density larger than that of the 
red points. Notice that with the exception of the CMSSM case at 
$M_{higgs} \approx 115$ GeV region, for a higgs mass above this value  
one can easily  saturate  the relic density. 
This is true in particular for points near the 125 GeV region, 
although in CMSSM this may become more problematic (too large
$\Omega h^2$).

The contour area of maximal values of $\delta a_\mu$ already
shown in previous figures 1-4, 17 is also presented. However, it should be stressed
that points inside this contour area that have the dark matter abundance as shown
(in blue, red or yellow) {\it are not necessarily the same points} that also have the
largest $\delta a_\mu$ within $2\sigma$  of $\delta a_\mu^{exp}$! (these are
different projections on the 2D plane shown). However points
that satisfy a relic density constraint {\it and} also have the largest $\delta a_\mu$
within $2\sigma$ of experimental value do fall within a smaller area inside
the contour shown.
 Of all models, NUGM could both saturate the relic density and 
fall within the  $\delta a_\mu$ contour line, for  large range of higgs mass, 
although the fine tuning cost grows exponentially
with $M_{higgs}$.
 Figures 5a), b) and 8 a), b) show again that
CMSSM and NUGM are the least sensitive models to any experimental constraints other 
than $\delta a_\mu$, for  $M_{higgs}$ larger than $\approx 115$ GeV (negligible
grey areas).  Finally, since $\Delta_q$, $\Delta_{max}$ are so similar, below we 
shall present only results for $\Delta_q$.

\subsection{$\Delta$ versus $M_{gluino}$ and $\Delta$ versus $m_{susy}$.}

So far we investigated the fine tuning as a function of the higgs mass. However, 
it is useful to present its dependence  on other particles 
masses, and we do this for  the gluino and the SUSY scale $m_{susy}$. 
This is useful since LHC searches for gluino or other SUSY partners can have a 
 strong impact on fine tuning. 
This is seen in figures~9-12 a), b), and figures~19 a), b),
  where  we show the dependence of $\Delta_q$
on the gluino mass (figures a))  and on $m_{susy}$ (figures~b))
 for all models. The light and dark grey areas have the 
same meaning as before, while the areas in black are ruled out by the higgs mass
constraint $111.4\leq M_{higgs}\leq 130$ GeV that we imposed
(this allows 2-3 GeV uncertainty for $M_{higgs}$
at two-loop leading log level \cite{Allanach:2001kg,PS,SH}).
Contour (dotted) lines of a {\it maximal} value of 123 GeV of $M_{higgs}$
are displayed for all models: the points below this line
respect this bound while those above can have larger values.  
The advantage of these plots is that  if future
data rules out $M_{higgs} <123$ GeV, the whole region below (outside) the dotted
line (contour)  will be  removed from the plots, to leave a small, restrictive
region.

$\delta a_\mu$ is also shown in colour 
encoded areas, with a red island area showing the largest possible value
with $\delta a_\mu^{max}$ within $2\sigma$ of the experimental central value. 
Note again that the  $\delta a_\mu$ contour and the dotted line of upper bounds 
on higgs mass are different projections on the 2D plane of the figures. That means
that  points that have largest $\delta a_\mu$ within $2\sigma$ of $\delta a_\mu^{exp}$
are not necessarily the same points that {\it simultaneously}
have  $M_{higgs}$ as large as 123 GeV.
The impact of future gluino mass or $m_{susy}$  bounds
from the LHC can easily be seen on these plots, together with the
associated fine tuning cost. The models
NUGM and NUGMd relax the lowest bound on the gluino mass due to 
their non-universal gaugino masses.

\begin{figure}[t!]\begin{center}
\includegraphics[width=14.cm,height=4.8cm]{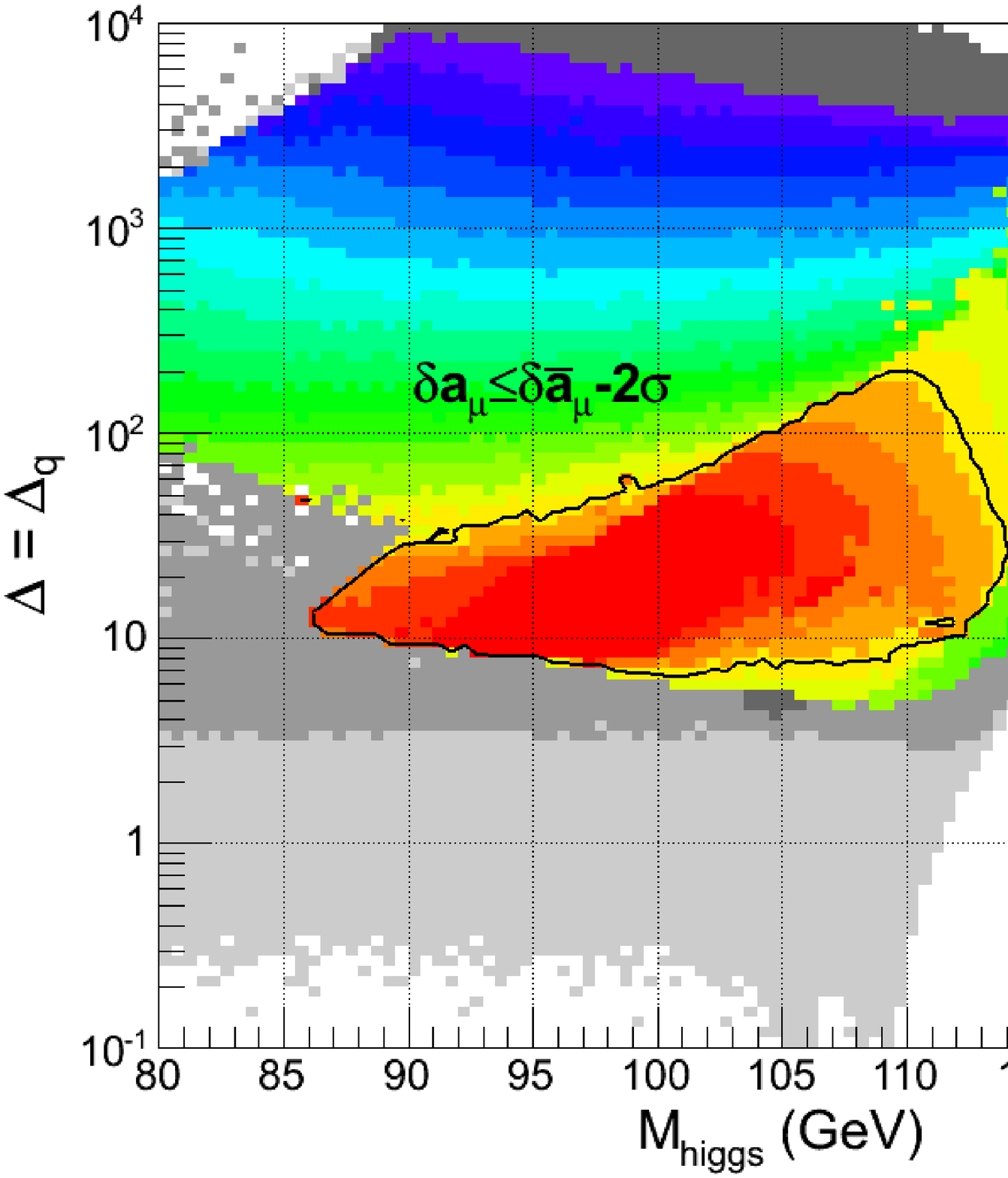}
\includegraphics[width=14.cm,height=4.8cm]{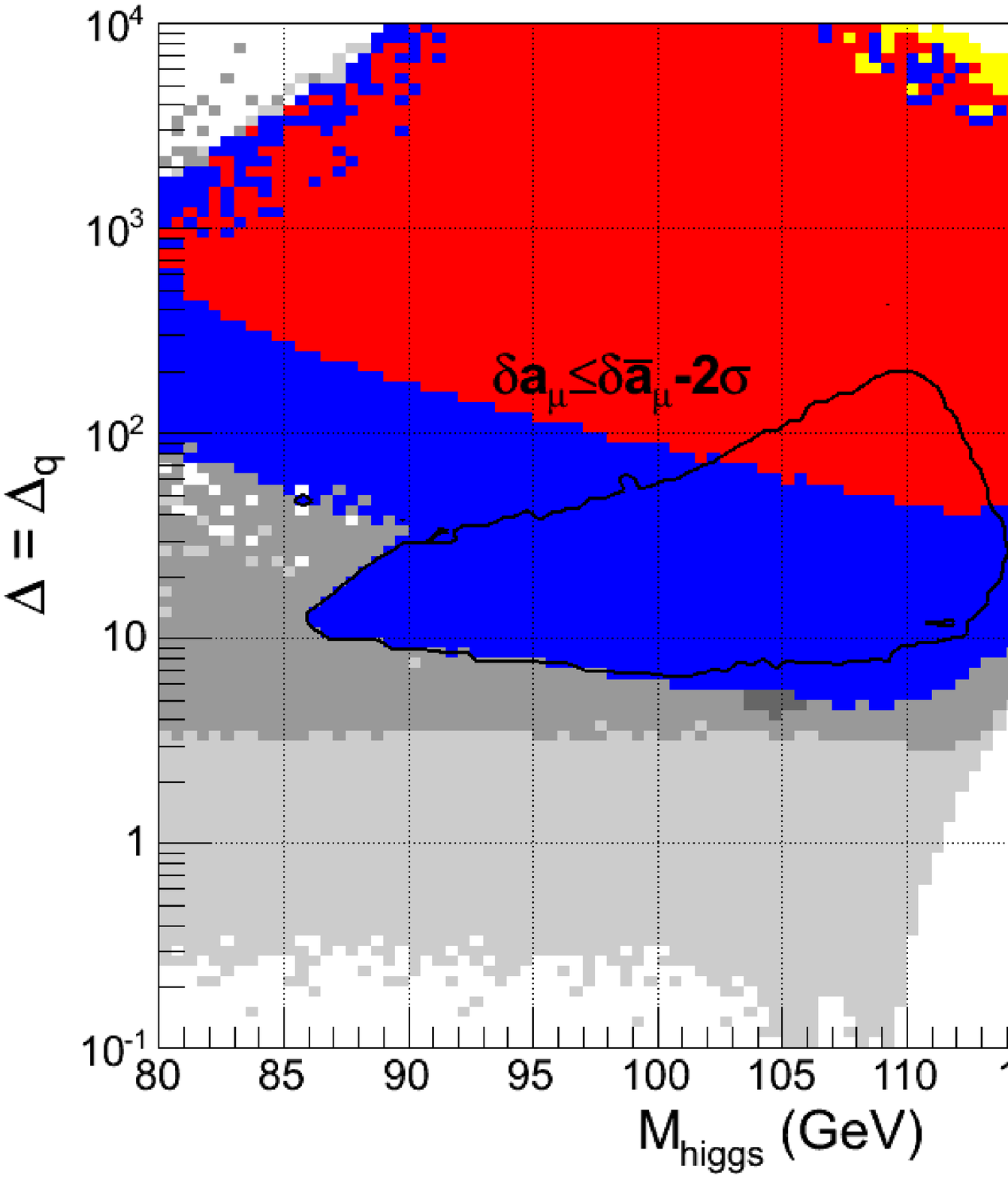}
\includegraphics[width=14.cm,height=4.8cm]{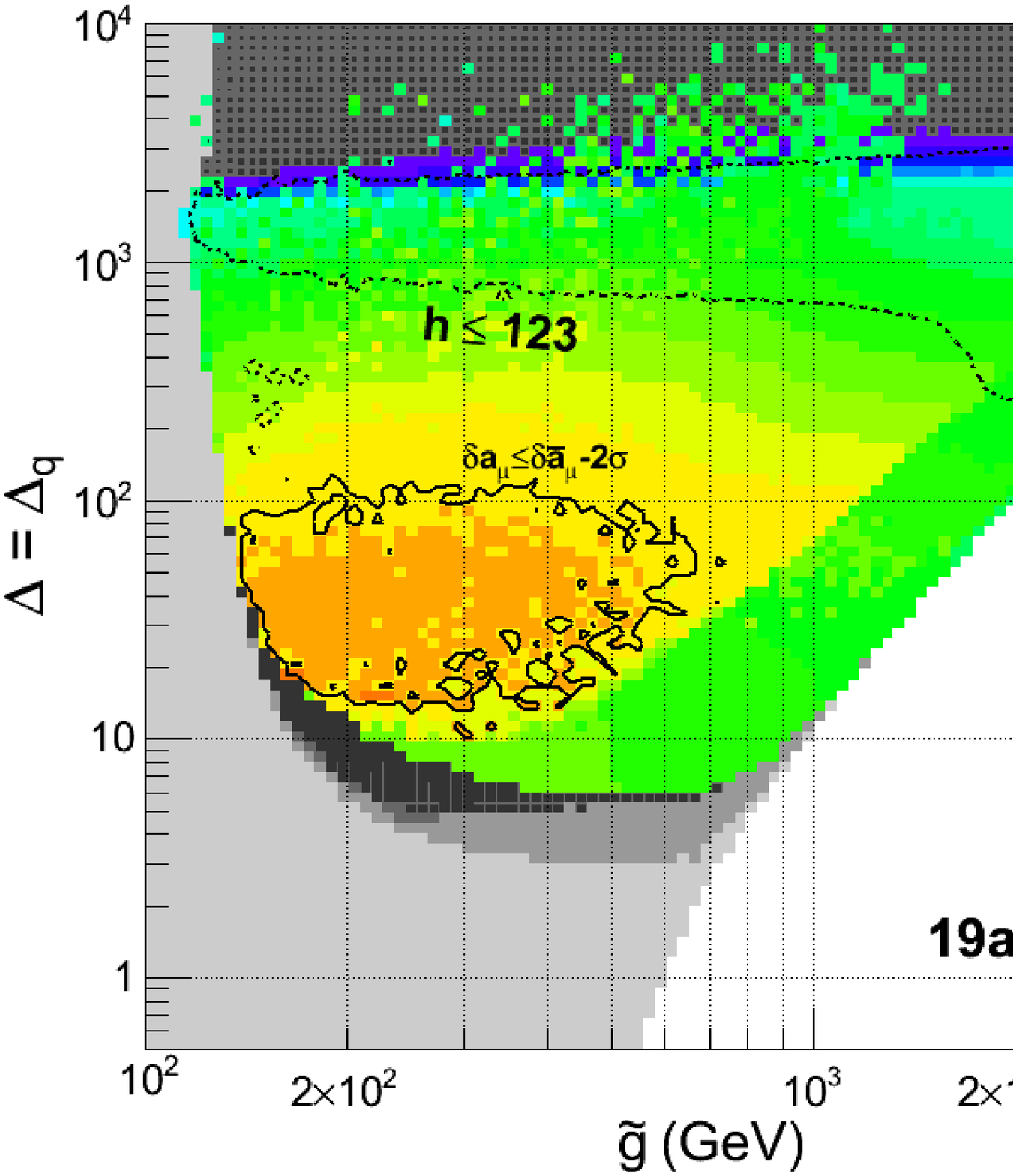}
\includegraphics[width=14.cm,height=4.8cm]{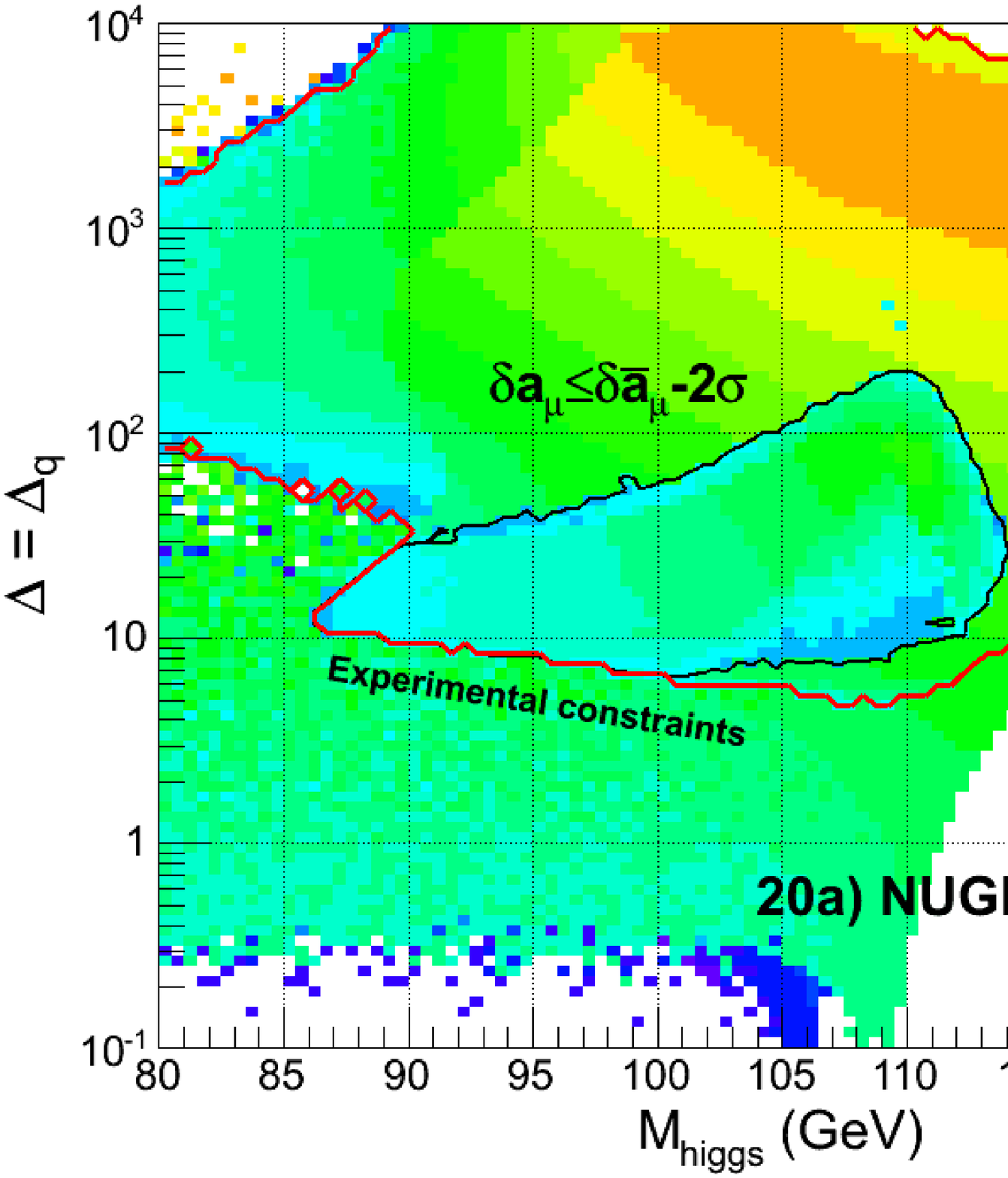}
\renewcommand{\baselinestretch}{0.9}
\caption*{\small 
Figures 17-20: The benchmark NUGMd model: 
the description of the plots  is identical to that in Figs. 1-17, 
but applied to  NUGMd model, as follows: 
Figs 17\,a),\,b) - as for figure 1\,a),\,b). 
Figs 18\,a),\,b) - as for figure 2\,a),\,b).
Figs 19\,a),\,b) - as for figure 3\,a),\,b). 
Figs 20\,a),\,b) - as for figure 4\,a),\,b).}
\label{figure11720}\end{center}\end{figure}

\subsection{$\Delta$ versus $M_{higgs}$ and the gluino mass range.}

A complementary presentation of  the results of figures 1-4 a), 17 a) and figures 9-12 a),
19 a), b) is that of figures 13-16 a), b) 
and 20 a), b). In these  $\Delta_q$ is presented again
as a function of the higgs mass,  but with the gluino mass as a parameter, with its
{\it largest} value in plots a) and {\it lowest} possible value in plots b), see the colour
encoded scale. In this
way one has a clear picture of the whole range of allowed values of gluino mass for a given 
$\Delta$ and higgs mass. Intermediate values of gluino mass are colour encoded. 
For the  large $M_{higgs}$, above 125 GeV the gluino mass tends to be larger (above 1 TeV), 
and within a narrow range, with increasing fine tuning cost. The range of values 
of gluino is rather similar in CMSSM, NUGM or in NUHM1, NUHM2.

One important remark about the contour of largest $\delta a_\mu$ shown: the 
gluino mass range shown inside this contour respects all experimental constraints,
{\it including} the constraint of $\delta a_\mu$ (within $2\sigma$)! 
If this constraint is lifted, the range of gluino mass, for a fixed higgs mass 
and fine tuning, 
would be larger. This also explains the sudden change of colour/spectra of gluino masses
around the contour line of $\delta a_\mu$ as compared to region immediately 
outside the contour.

\subsection{Stop versus gluino mass, with the largest  $M_{higgs}$ and 
minimal  $\Delta$.}

For a future comparison with results from LHC searches for new physics, we also present
in figures 21, the dependence stop versus gluino mass 
and with the {\it largest} value of Higgs mass that is
possible with the former two fixed. The {\it minimal} fine tuning cost that comes with this
is also shown in the corresponding areas (bordered by red contour lines),
while the largest $M_{higgs}$ allowed is colour encoded, see the 
scale on the right side of the plots.
The latest bounds on the gluino and stop masses can be translated into (upper) bounds
for the higgs mass. Currently,  stop-gluino exclusion plots from the LHC exist 
only for simple models that cannot be used for comparison~\cite{atlasgluinostop},
see the first plot in  figure 21. 
Eventually, at very large gluino and stop masses the (minimal)
fine tuning cost becomes too large and the models may be considered unrealistic.
It can be seen from these plots how the lowest allowed fine tuning increases as the
higgs mass goes towards its upper limit.
If one rules out values of fine tuning of say $\Delta_q\geq 100$ one immediately 
removes the area outside the contour line that borders this region, to leave a significantly
smaller area of correlation stop-gluino-higgs mass. 

 While the  CMSSM, NUHM1, NUHM2 are
more restricted by superpartners masses (excluded light grey areas), in the case of the 
NUGM and NUGMd, not surprisingly, the impact of the spartners mass bounds 
is small (since the universality condition was relaxed). 
Again, the NUGM model is less restricted, allowing a large higgs
mass (125-128 GeV), with a stop as light as $400-500$ GeV and gluino
mass between $2-3$ TeV.

\begin{figure}[t!]
\begin{center}
\psfig{figure=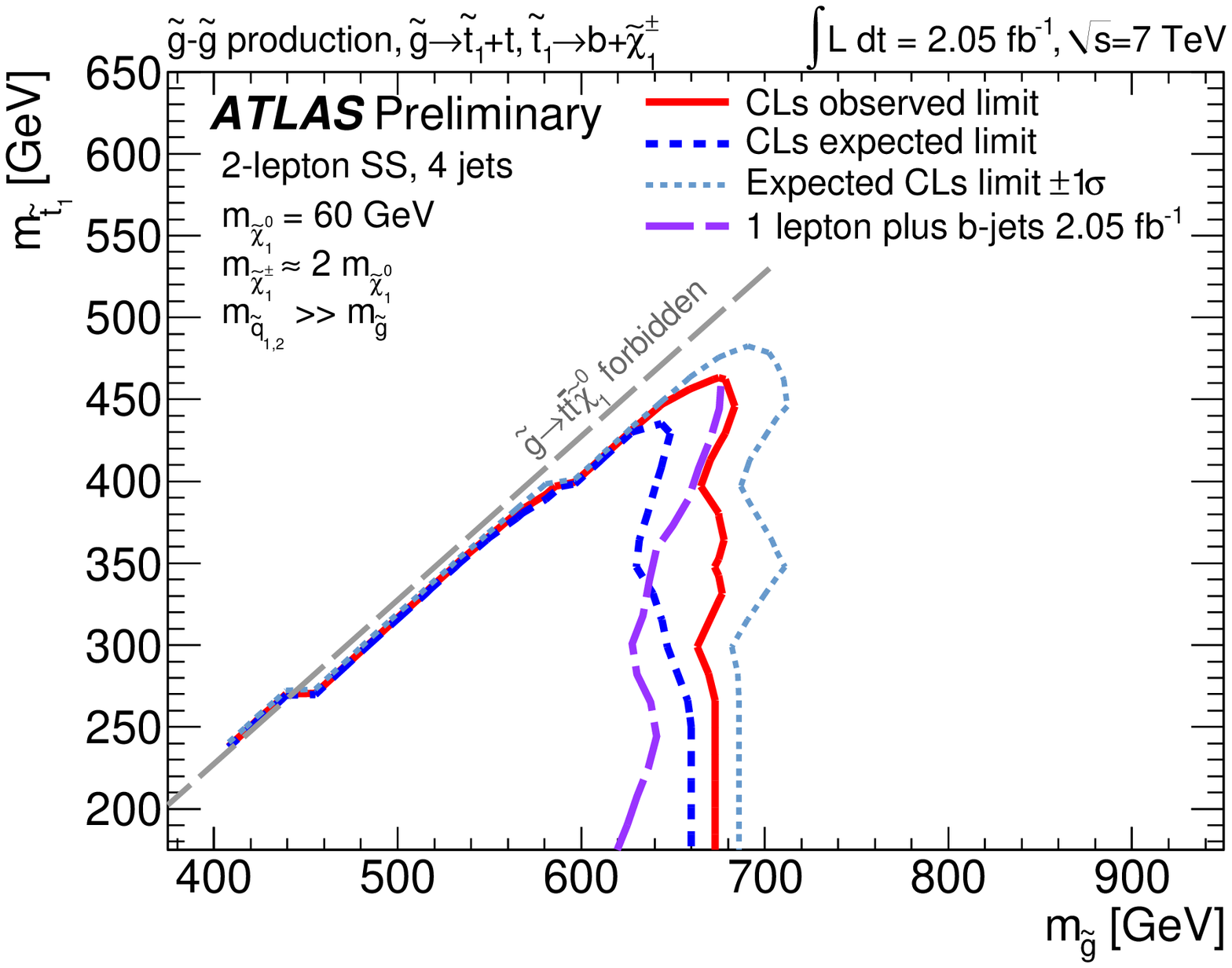,width=6.7cm,height=5.cm}
\hspace{0.cm}
\psfig{figure=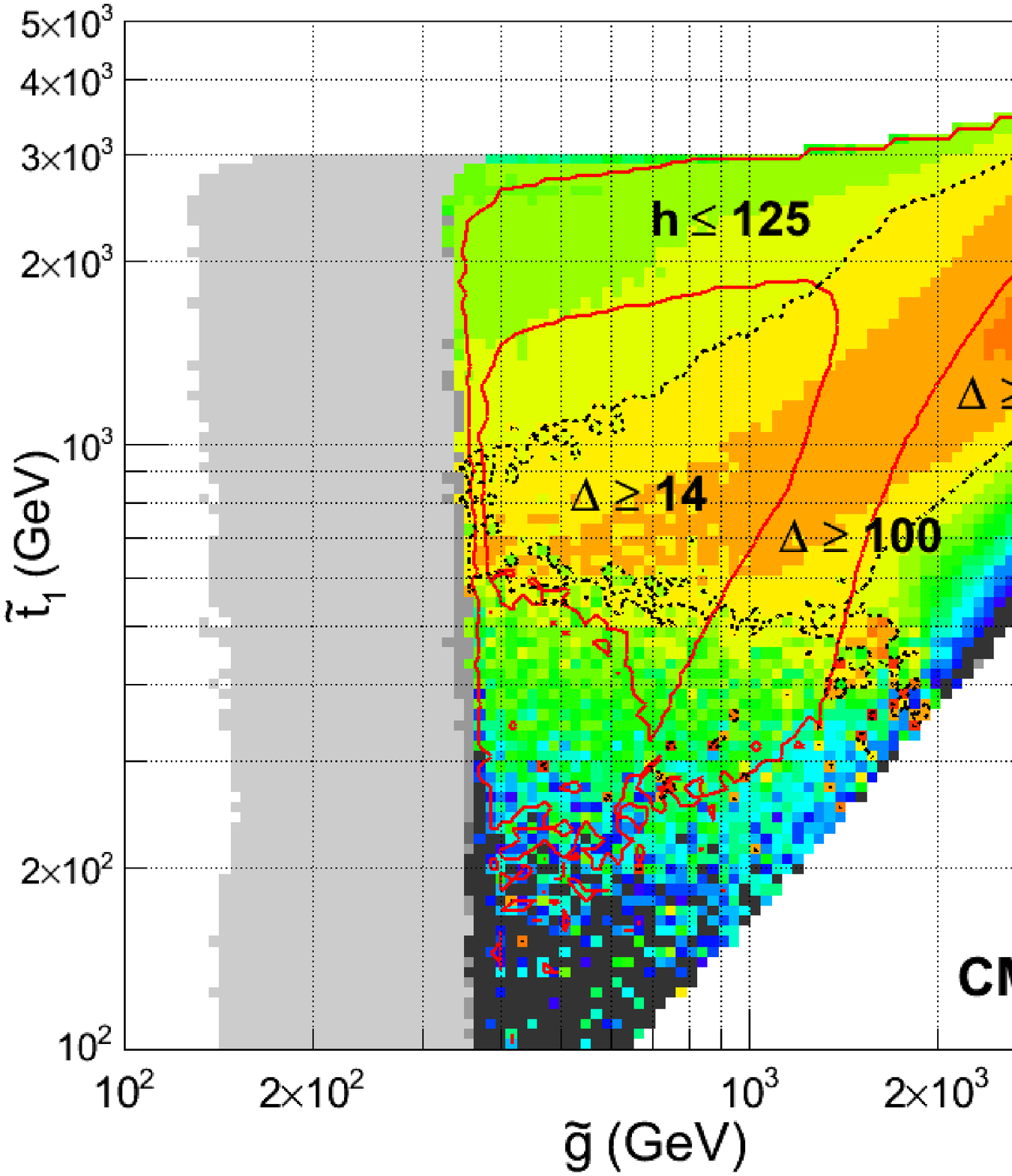,width=6.2cm,height=5.2cm}
\epsfig{figure=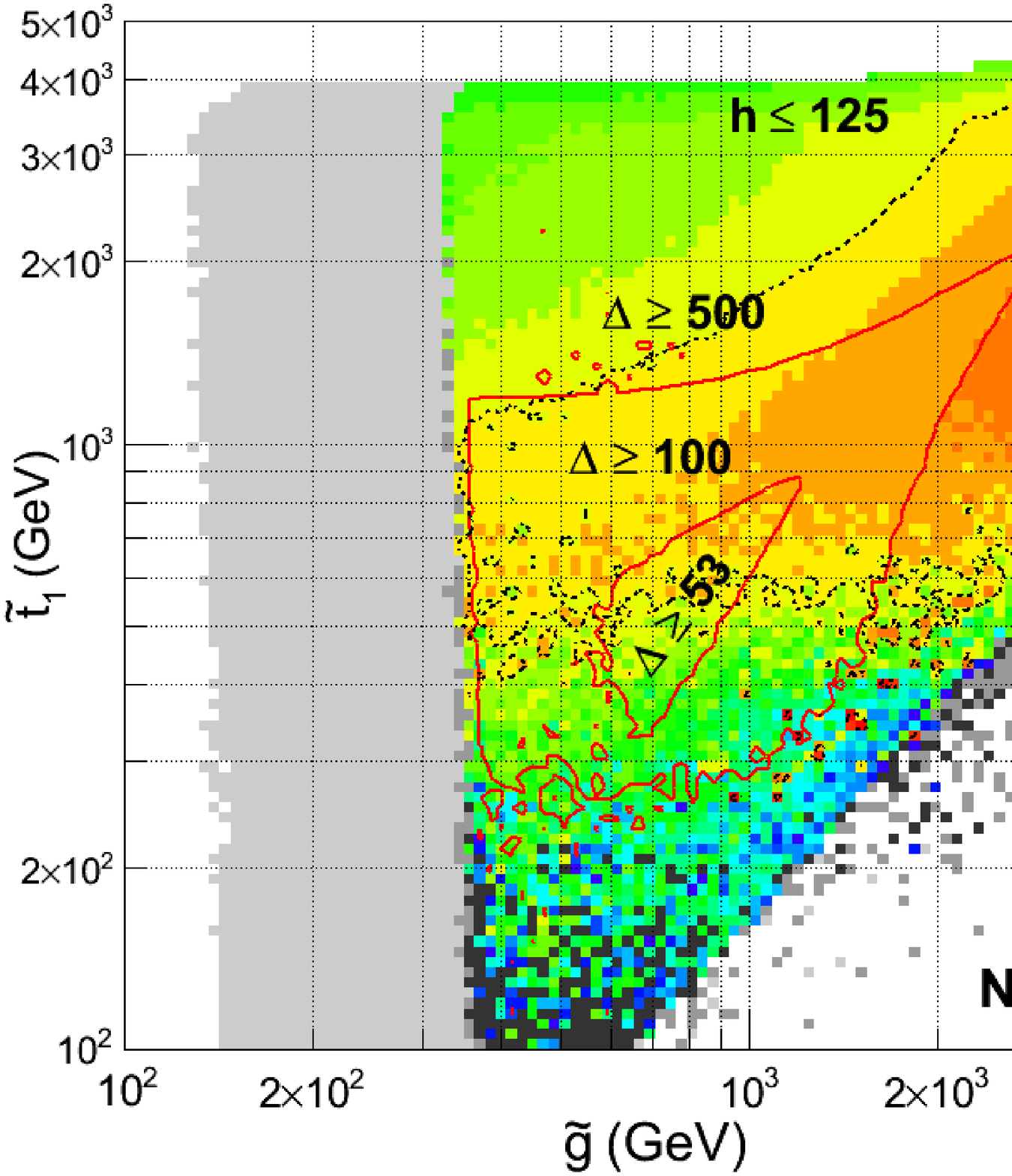,width=6.2cm,height=5.2cm}
\hspace{0.5cm}
\epsfig{figure=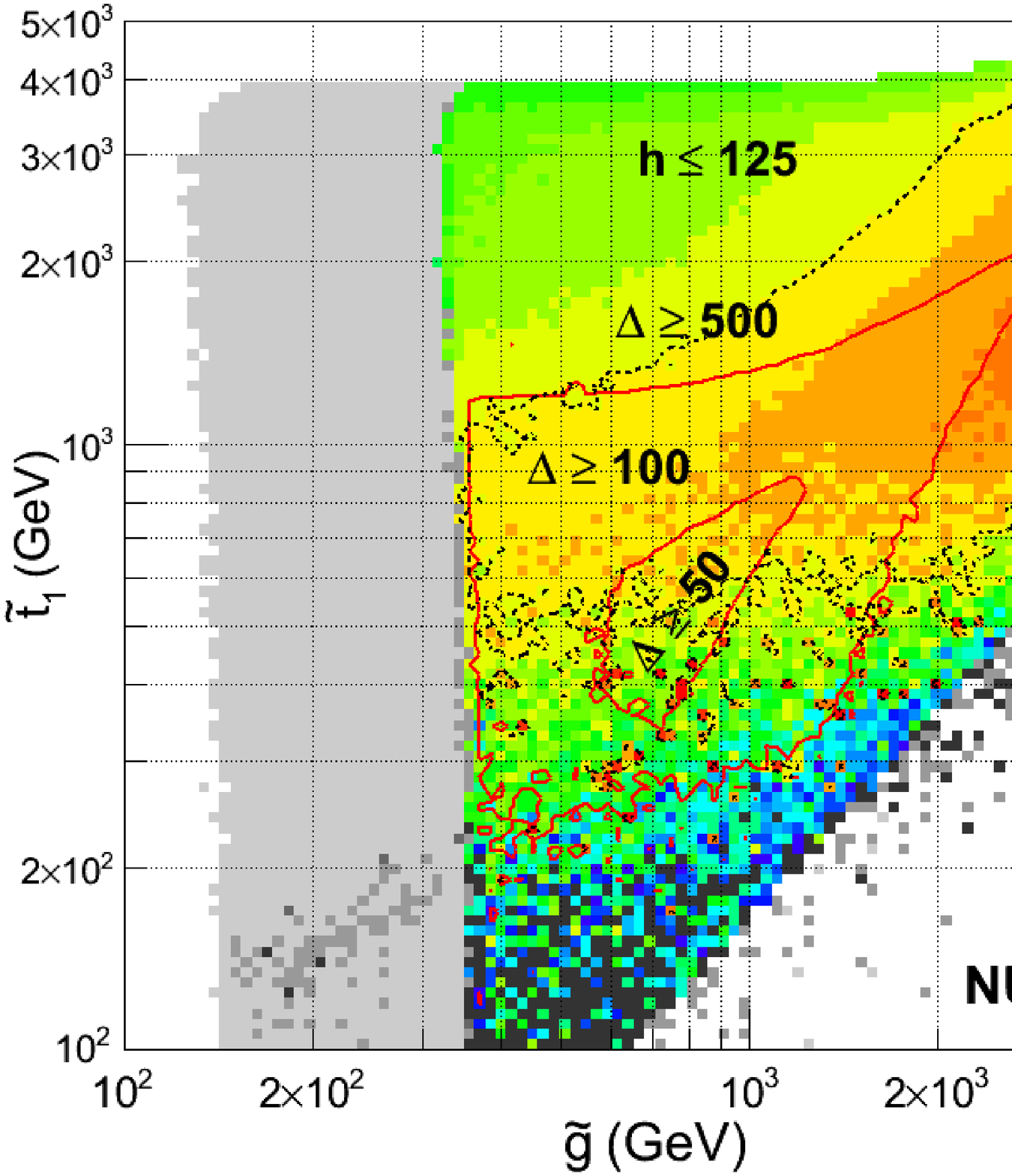,width=6.2cm,height=5.2cm}
\bigskip
\epsfig{figure=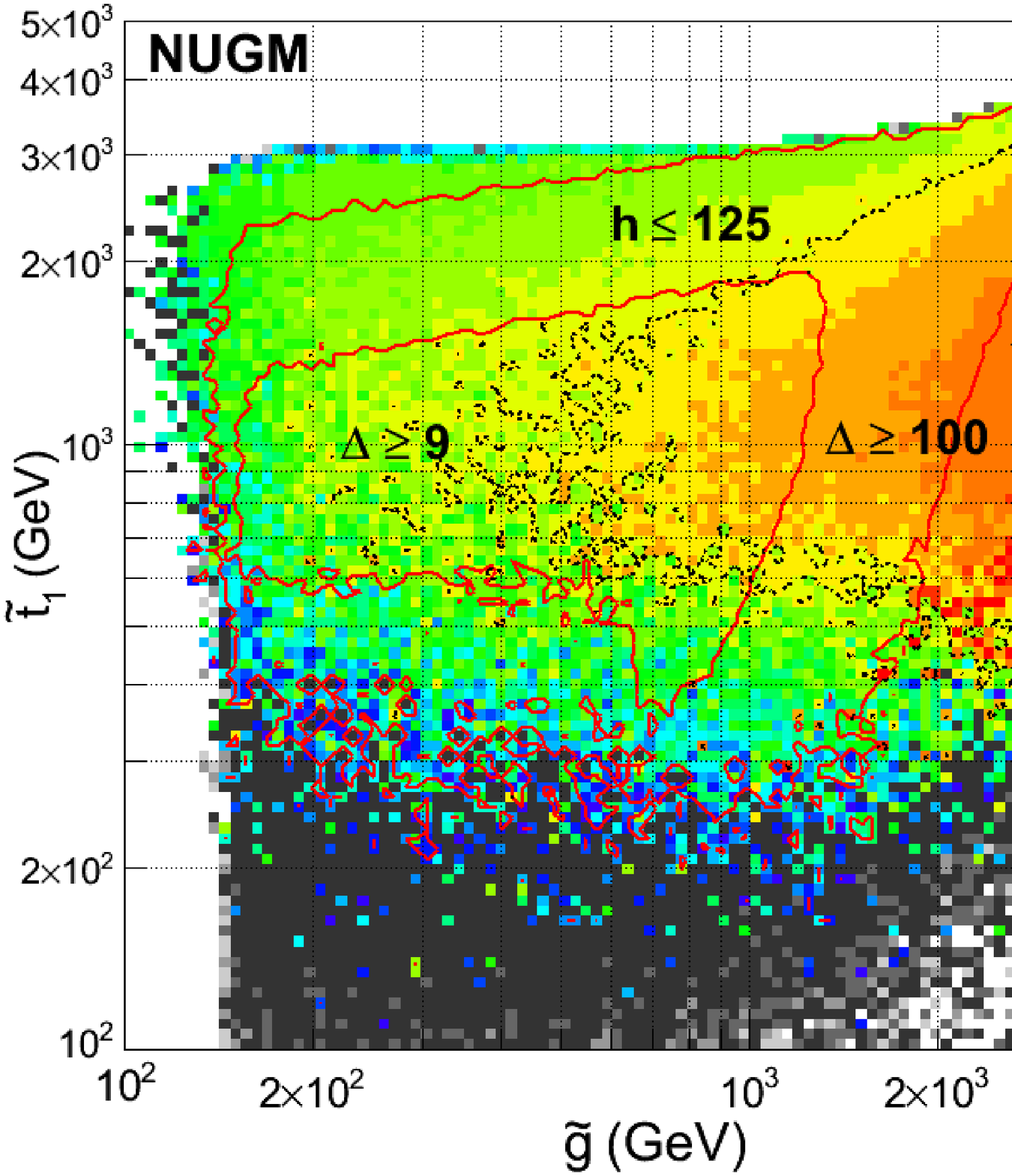,width=6.2cm,height=5.2cm}
\hspace{0.5cm}
\epsfig{figure=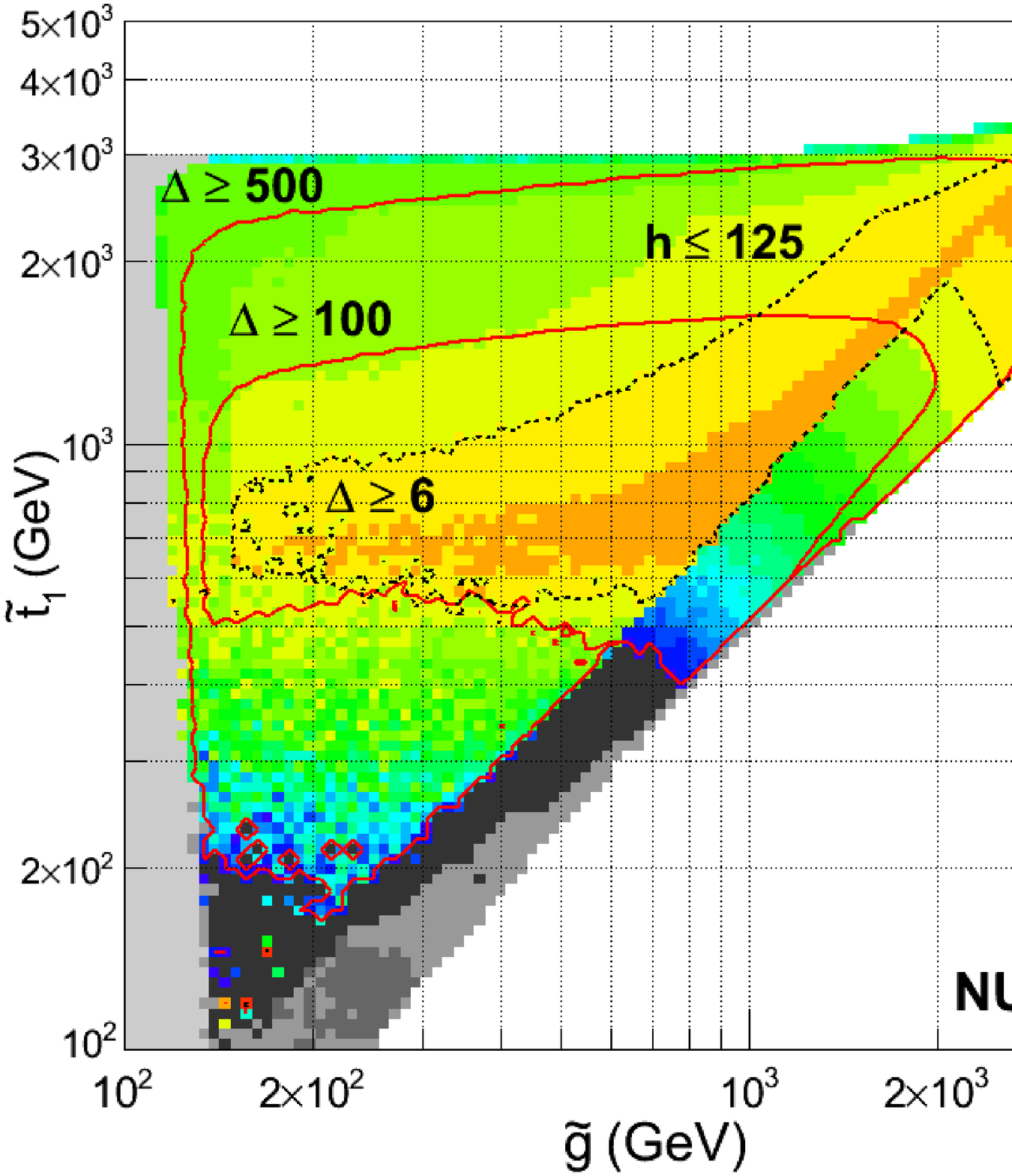,width=6.2cm,height=5.2cm}
\renewcommand{\baselinestretch}{0.9}
\caption*{\small Figure 21:
Top left plot: Atlas stop-gluino exclusion limits in a simple
supersymmetric model \cite{atlasgluinostop}.\newline
Rest of the plots:
The dependence stop vs gluino mass in CMSSM, NUHM1, NUHM2, NUGM, NUGMd
models, in this order. We present the {\it lowest} value of $\Delta_q$ in 
the  areas where it is shown,  bounded by red contour lines and 
with no upper bound.
Areas of {\it largest} Higgs mass are also shown, colour encoded,
see the scale on the right side (minimal value: 111.4 GeV). 
One can easily  see the {\it largest} higgs mass and the 
{\it minimal} fine tuning cost,  for given gluino and stop masses.
Grey area is excluded by SUSY mass bounds. Black area is excluded by
imposing the constraint $M_{higgs}\geq 111.4$ GeV and $M_{higgs}\leq 130$ GeV.
 This dependence can eventually be compared
with similar future  plots from CMS/ATLAS searches.}
\label{figurel}\end{center}
\end{figure}

\section{Final remarks and conclusions}

Low energy (TeV scale) supersymmetry is thought
to  solve the hierarchy problem without undue amount of 
fine tuning ($\Delta$). However there are 
different opinions on what the best definition for $\Delta$ 
is, or  what upper value is allowed for it  while still claiming a SUSY
solution to this problem. To avoid a subjective choice on these 
two issues, we performed a study of $\Delta$ using two common 
definitions  $\Delta_{max}$ and $\Delta_q$  and made no assumption about their largest
allowed values. We also discussed the relation of $\Delta$ to global 
probabilities (in the parameter space) to fit the  data.
We analyzed generic  models: CMSSM, NUHM1, NUHM2, 
NUGM and a benchmark model, NUGMd, at two-loop leading log level, and 
both $\Delta_{max}$ and $\Delta_q$ were  presented as functions of the higgs, 
gluino, stop mass or the SUSY scale, with additional constraints like
dark matter or $\delta a_\mu$.  The advantage of this comparative analysis is that 
using the figures for $\Delta_q$, $\Delta_{max}$, future experimental constraints 
can  immediately  be converted  into  an updated estimate for the fine 
tuning level of these models, without the need to re-do the whole analysis.  
The reader will then decide  whether  the amount of tuning so obtained 
is still acceptable for a solution to the hierarchy problem.

The measures of fine tuning were originally introduced more on physical intuition than
rigorous  mathematical grounds.
 {In this work we provided  mathematical support for the fine tuning via 
a quantitative relation to Bayesian evidence $p(D)$.}
 As   direct result of two theoretical  constraints
(EW min conditions), we showed that a  fine tuning measure $\tilde\Delta_q\!=\!\Delta_q$
emerges as an additional suppression factor 
(effective prior) of the averaged likelihood under the initial priors, under the 
integral of global probability of measuring the data (the evidence $p(D)$).
{So the Bayesian evidence calculation prefers  $\Delta_q$ as a fine tuning measure.}
 As a result, the  evidence $p(D)\sim 1/ \Delta_q$ therefore points of  large $\Delta_q$
(strongly fine tuned) have little or no impact on the global probability 
of the  model to fit the data.   {These results provide} technical support to
 the idea that fine tuning has 
a physical meaning and that preferably it should have small values
 in realistic models  {for the corresponding point in the parameter space.}

Our numerical results for  $\Delta_q$ and $\Delta_{max}$ as functions of the higgs mass, 
showed that they  {have close values for the same higgs mass} 
and also very similar {\it behaviour} for all models 
considered. There is a small discrepancy factor between them  {(between 1 and 2)}
which is most visible for regions of the higgs mass that 
are anyway excluded by the data. 
  {All these results show a good} independence on the actual definition used
for fine tuning.
For $115\leq M_{higgs}\leq 128$ GeV there is a relative 
independence of $\Delta_q$ or $\Delta_{max}$ on the experimental constraints 
(other than $\delta a_\mu$) for CMSSM or NUGM, NUGMd,  with a minor 
dependence for NUHM1, NUHM2. So in this case  $\Delta_q$, $\Delta_{max}$ 
 are largely controlled by  theoretical constraints.
Also, the dark matter relic density can in all cases be saturated
 within  {$3\sigma$} of the current value.

The dependence of both $\Delta$ on the gluino mass or on the SUSY scale shows a 
similar behaviour for all models. The CMSSM, NUGM and NUGMd models show a lower amount of
fine tuning for the same experimental constraints, and NUGM can even accommodate
 $\delta a_\mu$ ($2\sigma$) and $M_{higgs}\approx 125$ GeV, 
however in this case there is always a fine tuning cost. 
 As our plots showed, for the CMSSM no fine tuning amount can reconcile 
$\delta a_\mu$ ($2\sigma$) values considered (i.e. contour at $2\sigma$ in the plots),
with a  $M_{higgs}> 120$ GeV region which is situated outside this contour. 
For a Higgs mass near $125$ GeV, the fine tuning is of order 
$\cO(1000)$ in all models other than NUGM, NUGMd where it is of order $\cO(500)$.
There is a strong (roughly exponential) variation of $\Delta$  with $M_{higgs}$. 
A reduction of 2 GeV of $M_{higgs}$ can bring down both $\Delta$'s to 
$\Delta\approx 200$ to $500$, depending on the model. 
For $M_{higgs}=115$ GeV, $\Delta_q\sim\Delta_{max}\approx 10$ to $100$
and in the CMSSM this $\Delta$ corresponds to a global minimum.
Finally, let us mention that the combined effect of a  {$1\sigma$}
increase of the top mass and  a similar reduction of the measured
strong coupling at EW scale can reduce the fine tuning for a 
given $M_{higgs}$ by a factor  near 2 or so in the CMSSM case 
\cite{Cassel:2010px}. 
Although we did not studied it here (due to long CPU runs), we expect 
similar effect for the other models. This is because Yukawa corrections help 
radiative EW breaking (reducing $\Delta$) while QCD corrections have the 
opposite effect in the loop diagrams.

Are the values of fine tuning that we found too large?
 {Based on previously agreed but highly subjective ``reasonable'' values
of $\Delta\sim 10-100$, the answer is probably affirmative.}
However, a  clear answer is 
difficult, largely because $\Delta$  depends $\approx$exponentially on the higgs 
mass, so any  {small} correction to it has a strong impact on $\Delta$. 
But comparing all models,
for the same experimental constraints, there seems to be a  preference for NUGM case
when also considering the $\delta a_\mu$ constraint.
We let  the reader to make his own opinion, 
based on the above results and figures and also on future LHC data 
(on gluino, higgs, stop and $m_{susy}$) whose updated impact on our $\Delta$ 
can easily be obtained. Also it should be kept 
in mind that very simple new physics beyond these 
SUSY models (like CMSSM with a gauge singlet with a TeV-scale SUSY mass term or a massive $U(1)^\prime$) 
can  lead to  a very acceptable  $\Delta\approx \cO(10)$ for a higgs mass as large 
as 130 GeV  {\cite{Cassel:2009ps}}. Further, subjective criteria also
 exist in other approaches that compare the 
probability of various models, such as those based on the Bayesian approach. 
 {Indeed, the} evidence $p(D)$ also has some dependence on the priors choice (flat, log, etc), 
until eventually more data can improve our knowledge of the models. 
We hope that the clear link between fine tuning $\Delta_q$ and
$p(D)$ that we established together with our  plots for both $\Delta$'s will 
provide the starting point of a more  detailed study.




\begin{thebibliography}{99}
\bibitem{Cassel:2011zd}
  S.~Cassel and D.~M.~Ghilencea,
 ``A Review of naturalness and dark matter prediction for the Higgs mass in MSSM and beyond,''
  Mod.\ Phys.\ Lett.\ A {\bf 27} (2012) 1230003
  [arXiv:1103.4793 [hep-ph]].

\bibitem{Ellis:1986yg}
  J.~R.~Ellis, K.~Enqvist, D.~V.~Nanopoulos and F.~Zwirner,
 ``Observables in Low-Energy Superstring Models,''
  Mod.\ Phys.\ Lett.\ A {\bf 1} (1986) 57.

\bibitem{Cabrera:2008tj}
  M.~E.~Cabrera, J.~A.~Casas and R.~Ruiz de Austri,
  ``Bayesian approach and Naturalness in MSSM analyses for the LHC,''
  JHEP {\bf 0903} (2009) 075
  [arXiv:0812.0536 [hep-ph]].
  M.~E.~Cabrera, J.~A.~Casas and R.~Ruiz d Austri,
  ``MSSM Forecast for the LHC,''
  JHEP {\bf 1005} (2010) 043
  [arXiv:0911.4686 [hep-ph]].

\bibitem{AbdusSalam:2009qd}
  S.~S.~AbdusSalam, B.~C.~Allanach, F.~Quevedo, F.~Feroz and M.~Hobson,
  ``Fitting the Phenomenological MSSM,''
  Phys.\ Rev.\ D {\bf 81} (2010) 095012
  [arXiv:0904.2548 [hep-ph]].
  B.~C.~Allanach, K.~Cranmer, C.~G.~Lester and A.~M.~Weber,
 ``Natural priors, CMSSM fits and LHC weather forecasts,''
  JHEP {\bf 0708} (2007) 023
  [arXiv:0705.0487 [hep-ph]].
  B.~C.~Allanach,
  ``Naturalness priors and fits to the constrained minimal supersymmetric standard model,''
  Phys.\ Lett.\ B {\bf 635} (2006) 123
  [hep-ph/0601089].

\bibitem{Cassel:2010px}
  S.~Cassel, D.~Ghilencea, G.~G.~Ross,
  ``Testing SUSY at the LHC: Electroweak and Dark matter fine tuning at two-loop order,''
  Nucl.\ Phys.\ B {\bf 835} (2010) 110
  [arXiv:1001.3884 [hep-ph]].
  ``Testing SUSY,''
  Phys.\ Lett.\ B {\bf 687} (2010) 214
  [arXiv:0911.1134 [hep-ph]].

\bibitem{nui}
J.O.  Berger, B. Liseo, R.L. Wolpert, ``Integrated likelihood methods for eliminating 
nuisance parameters'', Statistical Science 1999, vol.14, No.1, 1-28.

\bibitem{Barbieri:1998uv} 
P.~H.~Chankowski, J.~R.~Ellis and S.~Pokorski, 
``The fine-tuning price of LEP,'' Phys.\ Lett.\ B \textbf{423} (1998) 327 
[arXiv:hep-ph/9712234].
P.~H.~Chankowski, J.~R.~Ellis, M.~Olechowski and 
S.~Pokorski, ``Haggling over the fine-tuning price of LEP,'' Nucl.\ Phys.\ B  
\textbf{544} (1999) 39 [arXiv:hep-ph/9808275].
 G.~L.~Kane and S.~F.~King, ``Naturalness implications 
of LEP results,'' Phys.\ Lett.\ B \textbf{451} (1999) 113 
[arXiv:hep-ph/9810374].
  R.~Barbieri and A.~Strumia,
``What is the limit on the Higgs mass?,''
  Phys.\ Lett.\ B {\bf 462} (1999) 144
  [hep-ph/9905281].
R.~Barbieri and A.~Strumia, ``About the 
fine-tuning price of LEP,'' 
Phys.\ Lett.\ B \textbf{433} (1998) 63 [arXiv:hep-ph/9801353].
  R.~Barbieri and G.~F.~Giudice,
``Upper Bounds on Supersymmetric Particle Masses,''
  Nucl.\ Phys.\ B {\bf 306} (1988) 63.

\bibitem{Horton:2009ed}
  D.~Horton and G.~G.~Ross,
  ``Naturalness and Focus Points with Non-Universal Gaugino Masses,''
  Nucl.\ Phys.\ B {\bf 830} (2010) 221
  [arXiv:0908.0857 [hep-ph]].

\bibitem{atlas2011}
ATLAS Collaboration,  F. Giannoti, 
``Update on the Standard Model  Higgs searches in
ATLAS,''. CERN Public Seminar, December 13, 2011, 
  G.~Aad {\it et al.}  [ATLAS Collaboration],
``Combined search for the Standard Model Higgs boson using
 up to 4.9 fb-1 of pp collision data at sqrt(s) = 7 TeV with the ATLAS detector at the LHC,''
  arXiv:1202.1408 [hep-ex].
  [ATLAS Collaboration],
 ``Search for the Standard Model Higgs boson in the diphoton decay
 channel with 4.9 fb-1 of pp collisions at sqrt(s)=7 TeV with ATLAS,''
  arXiv:1202.1414 [hep-ex].
  [ATLAS Collaboration],
``Search for the Standard Model Higgs boson in the decay channel
 H$\rightarrow$ ZZ(*)$\rightarrow$ 4l with 4.8 fb-1 of pp collisions at sqrt(s)=7 TeV with ATLAS,''
  arXiv:1202.1415 [hep-ex].

\bibitem{CMS2011}
CMS Collaboration, G. Tonelli, 
``Update on the Standard Model Higgs searches in CMS,''.
CERN Public Seminar, December 13, 2011,
  S.~Chatrchyan {\it et al.}  [CMS Collaboration],
``Search for the standard model Higgs boson decaying to 
bottom quarks in pp collisions at sqrt(s)=7 TeV,''
  arXiv:1202.4195 [hep-ex].
  S.~Chatrchyan {\it et al.}  [CMS Collaboration],
 ``Search for neutral Higgs bosons decaying to tau pairs in pp collisions at sqrt(s)=7 TeV,''
  arXiv:1202.4083 [hep-ex].
  S.~Chatrchyan {\it et al.}  [CMS Collaboration],
``Search for the standard model Higgs boson decaying into two
 photons in pp collisions at sqrt(s)=7 TeV,''
  arXiv:1202.1487 [hep-ex].
  S.~Chatrchyan {\it et al.}  [CMS Collaboration],
``Combined results of searches for the standard model Higgs
 boson in pp collisions at sqrt(s) = 7 TeV,''
  arXiv:1202.1488 [hep-ex].


\bibitem{Cassel:2009ps}
  S.~Cassel, D.~M.~Ghilencea and G.~G.~Ross,
  ``Fine tuning as an indication of physics beyond the MSSM,''
  Nucl.\ Phys.\ B {\bf 825} (2010) 203
  [arXiv:0903.1115 [hep-ph]].
  G.~G.~Ross, K.~Schmidt-Hoberg and F.~Staub,
  ``The generalised NMSSM at one loop: fine tuning and phenomenology,''
  arXiv:1205.1509 [hep-ph];
G.~G.~Ross and K.~Schmidt-Hoberg,
 ``The fine-tuning of the generalised NMSSM,''
  Nucl.\ Phys.\ B {\bf 862} (2012) 710.
  M.~Carena, K.~Kong, E.~Ponton and J.~Zurita,
  ``Supersymmetric Higgs Bosons and Beyond,''
  Phys.\ Rev.\ D {\bf 81} (2010) 015001
  [arXiv:0909.5434 [hep-ph]].
  I.~Antoniadis, E.~Dudas, D.~M.~Ghilencea and P.~Tziveloglou,
  ``MSSM Higgs with dimension-six operators,''
  Nucl.\ Phys.\ B {\bf 831} (2010) 133
  [arXiv:0910.1100 [hep-ph]].
  I.~Antoniadis, E.~Dudas, D.~M.~Ghilencea and P.~Tziveloglou,
  ``Beyond the MSSM Higgs with d=6 effective operators,''
  Nucl.\ Phys.\ B {\bf 848} (2011) 1
  [arXiv:1012.5310 [hep-ph]].
  I.~Antoniadis, E.~Dudas, D.~M.~Ghilencea and P.~Tziveloglou,
  ``Non-linear MSSM,''
  Nucl.\ Phys.\ B {\bf 841} (2010) 157
  [arXiv:1006.1662 [hep-ph]].
  M.~Carena, E.~Ponton and J.~Zurita,
  ``BMSSM Higgs Bosons at the Tevatron and the LHC,''
  Phys.\ Rev.\ D {\bf 82} (2010) 055025
  [arXiv:1005.4887 [hep-ph]].
  M.~Carena, E.~Ponton and J.~Zurita,
  ``BMSSM Higgs Bosons at the 7 TeV LHC,''
  Phys.\ Rev.\ D {\bf 85} (2012) 035007
  [arXiv:1111.2049 [hep-ph]].
  A.~Brignole, J.~A.~Casas, J.~R.~Espinosa and I.~Navarro,
  ``Low scale supersymmetry breaking: Effective description,
 electroweak breaking and phenomenology,''
  Nucl.\ Phys.\ B {\bf 666} (2003) 105
  [hep-ph/0301121].


\bibitem{Martin:1993zk}
  S.~P.~Martin and M.~T.~Vaughn,
  ``Two Loop Renormalization Group 
Equations For Soft Supersymmetry Breaking Couplings,''
  Phys.\ Rev.\  D {\bf 50}, 2282 (1994)
  [Erratum-ibid.\  D {\bf 78}, 039903 (2008)]
  [arXiv:hep-ph/9311340].

\bibitem{Carena:1995bx} 
M.~S.~Carena, J.~R.~Espinosa, M.~Quiros and C.~E.~M.~Wagner, 
``Analytical expressions for radiatively corrected Higgs 
 masses and couplings in the MSSM,''
Phys.\ Lett.\ B \textbf{355} (1995) 209 [arXiv:hep-ph/9504316].


\bibitem{Pythia8}
  T.~Sjostrand, S.~Mrenna, P.~Z.~Skands,
  ``A Brief Introduction to PYTHIA 8.1,''
  Comput.\ Phys.\ Commun.\  {\bf 178} (2008) 852
  [arXiv:0710.3820 [hep-ph]].
http://home.thep.lu.se/~torbjorn/pythiaaux/present.html

\bibitem{Belanger:2001fz}
G.~Belanger, F.~Boudjema, A.~Pukhov and A.~Semenov,
``MicrOMEGAs: A Program for calculating the relic density in the MSSM,''
 Comput.\ Phys.\ Commun.\  {\bf 149} (2002) 103 [hep-ph/0112278].
``micrOMEGAs: Version 1.3,''
Comput.\ Phys.\ Commun.\  {\bf 174} (2006) 577 [hep-ph/0405253].
 ``MicrOMEGAs 2.0: A Program to calculate the relic density of dark matter in 
a generic model,''  Comput.\ Phys.\ Commun.\  {\bf 176} (2007) 367 [hep-ph/0607059].

\bibitem{Allanach:2001kg}
  B.~C.~Allanach,
 ``SOFTSUSY: a program for calculating supersymmetric spectra,''
  Comput.\ Phys.\ Commun.\  {\bf 143} (2002) 305
  [hep-ph/0104145].

\bibitem{Buchmueller:2011sw}
  O.~Buchmueller, R.~Cavanaugh, A.~De Roeck, M.~J.~Dolan,
 J.~R.~Ellis, H.~Flacher, S.~Heinemeyer and G.~Isidori {\it et al.},
  ``Supersymmetry in Light of 1/fb of LHC Data,''
  arXiv:1110.3568 [hep-ph].
  S.~Sekmen, S.~Kraml, J.~Lykken, F.~Moortgat, S.~Padhi, L.~Pape,
 M.~Pierini and H.~B.~Prosper {\it et al.},
  ``Interpreting LHC SUSY searches in the phenomenological MSSM,''
  arXiv:1109.5119 [hep-ph].
S.~Cassel, D.~M.~Ghilencea, S.~Kraml, A.~Lessa, G.~G.~Ross,
``Fine-tuning implications for complementary dark matter and LHC SUSY searches,''
[arXiv:1101.4664 [hep-ph]].
  S.~Akula, N.~Chen, D.~Feldman, M.~Liu, Z.~Liu, P.~Nath and G.~Peim,
  ``Interpreting the First CMS and ATLAS SUSY Results,''
  arXiv:1103.1197 [hep-ph].
  D.~Feldman, K.~Freese, P.~Nath, B.~D.~Nelson and G.~Peim,
  ``Predictive Signatures of Supersymmetry: Measuring the Dark Matter Mass and
  Gluino Mass with Early LHC data,''
  arXiv:1102.2548 [hep-ph].
  P.~Bechtle, K.~Desch, H.~K.~Dreiner, M.~Kramer, B.~O'Leary, C.~Robens, B.~Sarrazin, P.~Wienemann,
  ``What if the LHC does not find supersymmetry in the sqrt(s)=7 TeV run?,''
   [arXiv:1102.4693 [hep-ph]].
  S.~Akula, D.~Feldman, Z.~Liu, P.~Nath, G.~Peim,
  ``New Constraints on Dark Matter from CMS and ATLAS Data,''
    [arXiv:1103.5061 [hep-ph]]

\bibitem{gminus2}
  M.~Davier, A.~Hoecker, B.~Malaescu, C.~Z.~Yuan and Z.~Zhang,
 ``Reevaluation of the hadronic contribution to the muon magnetic anomaly using 
new e+ e- $\rightarrow$ pi+ pi- cross section data from BABAR,''
  Eur.\ Phys.\ J.\ C {\bf 66} (2010) 1
  [arXiv:0908.4300 [hep-ph]].

\bibitem{bsg}
 D.~Asner {\it et al.}  [Heavy Flavor Averaging Group Collaboration],
 ``Averages of b-hadron, c-hadron, and $\tau$-lepton Properties,''
 arXiv:1010.1589 [hep-ex].

\bibitem{bmupmum}
The CMS and LHCb Collaborations, CMS-PAS-BPH-11-019, LHCb-CONF-2011-047,
CERN-LHCb-CONF-2011-047. 

%
\bibitem{pdg} The Particle Data Group, http://pdg.lbl.gov/

\bibitem{wmap}
  G.~Hinshaw {\it et al.}  [WMAP Collaboration],
 ``Five-Year Wilkinson Microwave Anisotropy Probe (WMAP) Observations:
 Data Processing, Sky Maps, and Basic Results,''
  Astrophys.\ J.\ Suppl.\  {\bf 180} (2009) 225
  [arXiv:0803.0732 [astro-ph]].

\bibitem{Aaij:2012ac}
  R.~Aaij {\it et al.}  [LHCb Collaboration],
  ``Strong constraints on the rare decays $B_s \to \mu^+ \mu^-$ and $B^0 \to \mu^+ \mu^-$,''
  arXiv:1203.4493 [hep-ex].
\bibitem{LEP2}
  S.~Schael {\it et al.}  [ALEPH and DELPHI and L3 and OPAL and LEP 
Working Group for Higgs Boson Searches Collaborations],
  ``Search for neutral MSSM Higgs bosons at LEP,''
  Eur.\ Phys.\ J.\ C {\bf 47} (2006) 547
  [hep-ex/0602042].
  R.~Barate {\it et al.},
  ``Search for the standard model Higgs boson at LEP,''
  Phys.\ Lett.\ B {\bf 565} (2003) 61
  [hep-ex/0306033].


\bibitem{PS}
G. Degrassi, S. Heinemeyer, W. Hollik, P. Slavich, and G. Weiglein, 
``Towards high precision predictions for the MSSM Higgs sector,'' 
Eur.Phys.J. C28 (2003) 133, arXiv:hep-ph/0212020 [hep-ph].

\bibitem{SH}
S. Heinemeyer, ``MSSM Higgs physics at higher orders,'' 
Int.J.Mod.Phys. A21 (2006) 2659, arXiv:hep-ph/0407244 [hep-ph].

\bibitem{atlasgluinostop}
The ATLAS collaboration, ATLAS-CONF-2012-004 
``Search for gluinos in events with two same-sign leptons,
 jets and missing transverse momentum with the ATLAS detector in $pp$ collisions at $\sqrt{s}=$7~TeV''
\end{thebibliography}
\end{document}